\documentclass[pre,floats,twocolumn,superscriptaddress,floatfix]{revtex4-1}

\usepackage{graphicx}
\usepackage{rotate}
\usepackage{epsfig}
\usepackage[usenames,dvipsnames]{xcolor}
\usepackage[normalem]{ulem}
\usepackage{amssymb,amsfonts,amsmath}
\usepackage{hyperref}
\usepackage{mathtools}
\usepackage{xparse} 
\usepackage{float}
\usepackage{colortbl}
\usepackage{multirow}

\definecolor{MyBrick}{rgb}{0.84,0.01,0.01}

\newcommand{\avg}[1]{\langle #1 \rangle}
\newcommand{\fc}{f_{\rm C}}

\newcommand{\fz}{f_{\rm Z}}
\newcommand{\pic}{\pi_{\rm C}}
\newcommand{\pid}{\pi_{\rm D}}

\newcommand{\etal}{\emph{et al.} }
\newcommand{\ie}{\emph{i.e.},}
\newcommand{\eg}{\emph{e.g.},}

\graphicspath{{pictures/}}

\makeatletter

    \def\CT@@do@color{%
      \global\let\CT@do@color\relax
            \@tempdima\wd\z@
            \advance\@tempdima\@tempdimb
            \advance\@tempdima\@tempdimc
    \advance\@tempdimb\tabcolsep
    \advance\@tempdimc\tabcolsep
    \advance\@tempdima2\tabcolsep
            \kern-\@tempdimb
            \leaders\vrule
                    \hskip\@tempdima\@plus  1fill
            \kern-\@tempdimc
            \hskip-\wd\z@ \@plus -1fill }
    \makeatother

\begin{document}

\title{Critical mass effect in evolutionary games triggered by zealots}

\author{Alessio Cardillo}
\affiliation{Department of Engineering Mathematics, University of Bristol, Bristol, BS8 1UB, United Kingdom}
\affiliation{Department of Computer Science and Mathematics, University Rovira i Virgili, E-43007 Tarragona, Spain}
\affiliation{GOTHAM Lab -- Institute for Biocomputation and Physics of Complex Systems (BIFI), University of Zaragoza, E-50018 Zaragoza, Spain}

\author{Naoki Masuda}
\email{naokimas@buffalo.edu}
\affiliation{Department of Mathematics, State University of New York at Buffalo, Buffalo, New York 14260-2900, USA}
\affiliation{Computational and Data-Enabled Science and Engineering Program, State University of New York at Buffalo, Buffalo, New York 14260-5030, USA}
\affiliation{Department of Engineering Mathematics, University of Bristol, Bristol, BS8 1UB, United Kingdom}

\begin{abstract}
Tiny perturbations may trigger large responses in systems near criticality, shifting them across equilibria. Committed minorities are suggested to be responsible for the emergence of collective behaviors in many physical, social, and biological systems. Using evolutionary game theory, we address the question whether a finite fraction of zealots can drive the system to large-scale coordination. We find that a tipping point exists in coordination games, whereas the same phenomenon depends on the selection pressure, update rule, and network structure in other types of games. Our study paves the way to understand social systems driven by the individuals' benefit in presence of zealots, such as human vaccination behavior or cooperative transports in animal groups.
\end{abstract}

\maketitle


\section{Introduction}
\label{sec:intro}

One hallmark of complex systems at the critical point is that small perturbations may trigger large responses, shifting the system from one equilibrium to another \cite{eckmann-rev_mod_phys-1985,buldyrev-nature-2010,gleeson-prx-2016,pound-j_am_chem_soc-1952,achlioptas-science-2009,d_souza-adv_phys-2019}. Small perturbations can be responsible for the emergence/disruption of collective phenomena such as synchronization \cite{kori-prl-2004,brocard-neuron-2013}, active Brownian motion (flocking) \cite{yllanes-njp-2017}, and cultural evolution \cite{young-j_econ_soc-1993,juul-pre-2019}. One way to model perturbations is to assume the existence of a \emph{committed minority}, whose fluctuations may trigger a system-wide response. Committed minorities also play a pivotal role in the emergence of \emph{consensus} among opinions and in coordination problems \cite{baronchelli-r_soc_op_sci-2018,galam-phys_a-2007}. Over the time, committed minorities have spurred a cascade of behavioral changes leading to a shift in the conventions held by the majority of the population (\eg{} civil rights movements \cite{tyson-j_am_hist-1998}, riots and revolutions \cite{moss_kanter-am_jour_soc-1977,opp-am_soc_rev-1993}, and vaccine hesitancy \cite{rochmyaningsih-science-2018}).

We refer to a committed individual, \ie{} having strong beliefs about something, as a \emph{zealot}. Zealotry can elicit consensus in human/animal behavior \cite{dyer-animal_beh-2008,couzin-science-2011,feinerman-nphys-2018}, the polarization of opinions in the voter model \cite{mobilia-prl-2003}, majority rule \cite{galam-phys_a-2007}, naming game \cite{xie-pre-2011,mistry-pre-2015,waagen-pre-2015}, social knowledge strucure (SKS) \cite{rodriguez-pone-2016}, and cooperative decision making model (CDMM) \cite{turalska-scirep-2013} models. Zealotry can also drive the emergence of cooperation in evolutionary games \cite{masuda-scirep-2012,mobilia-pre-2012,nakajima-math_bio-2015}, and the attainment of the optimal equilibrium in the Schelling's model of social segregation dynamics \cite{jensen-prl-2018}. Conversely, zealous (\ie{} stubborn) dissenters can disrupt flocking in the Vicsek's active matter model \cite{yllanes-njp-2017}. Lastly, including zealots of opposite types hinders opinions' polarization in the voter model \cite{mobilia-jstat-2007} and majority rule \cite{galam-phys_a-2007}. One question about zealots is whether or not their catalytic role in migrating the system across equilibria requires a \emph{critical mass}. Specifically, an infinitesimal fraction of zealots is enough to shift the equilibria in the voter \cite{mobilia-prl-2003}, CDMM \cite{turalska-scirep-2013}, and Schelling's \cite{jensen-prl-2018} dynamics, whereas finite fractions are needed in both the majority rule \cite{galam-phys_a-2007} and Vicsek \cite{yllanes-njp-2017} dynamics, as well as in the naming game theoretically \cite{xie-pre-2011,waagen-pre-2015} and experimentally \cite{centola-science-2018}.

In this paper, we clarify the presence of critical mass effects induced by zealots by providing a comprehensive study based on evolutionary game theory. We examine the conditions under which a critical mass of zealots enabling the consensus of one opinion state, which one often identifies with cooperation in the context of social dilemma games, exists. We extend the results for well-mixed populations presented in Refs.~\cite{masuda-scirep-2012,nakajima-math_bio-2015} by considering a wider class of games, distinct update rules, and making a thorough analysis of the effects of selection pressure on the critical mass effect. We provide a wider overview on the overall phenomenology and a connection with analogous phenomena observed in opinion dynamics. Contrary to the assumption made in Ref.~\cite{mobilia-pre-2012}, our zealous agents contribute to the payoff of any agent, regardless of its opinion state. We also consider the case of networks of agents and investigate the role of network structure on the appearance of the critical mass effect.

%
%
%
%
%

\section{Model}
\label{sec:model}

\begin{figure}[t!]
\centering
\includegraphics[width=0.45\columnwidth]{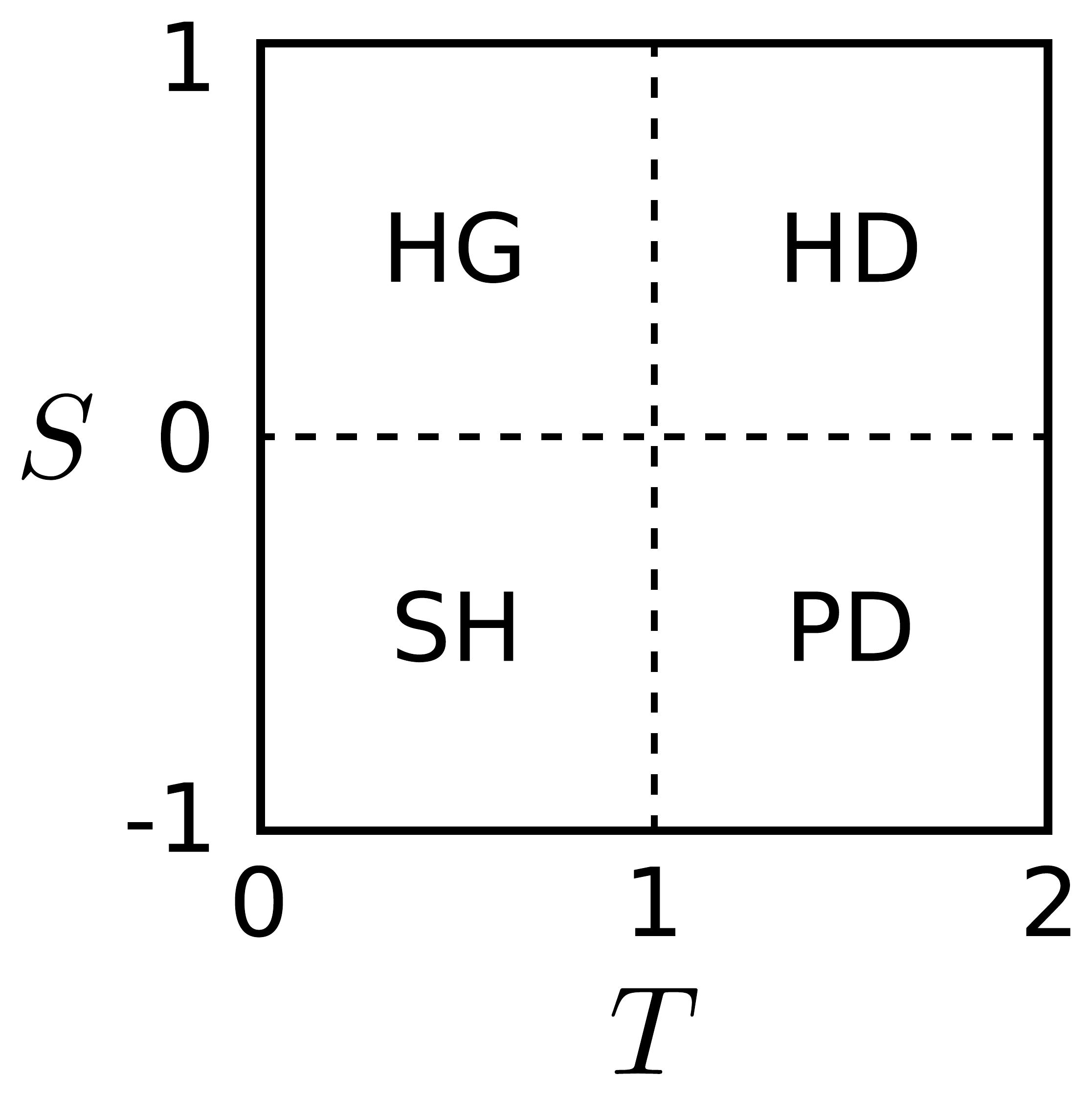}
\caption{Schematic representation of the four possible dilemmas available in the $(T,S)$ space according to payoff matrix $\mathcal{A}$. We have: the Harmony Game (HG), the Hawk and Dove (HD), the Prisoner's Dilemma (PD), and the Stag Hunt (SH).}
\label{fig:games-scheme}
\end{figure}

Let us consider a well-mixed population of $N$ agents playing an evolutionary two-strategy game. The state of an agent $i$, $1 \leq i \leq N$, is determined by its \emph{strategy}, $s_i$, which we refer to as either \emph{cooperation} ($s_i = {\rm C}$) or \emph{defection} ($s_i = {\rm D}$). The entries of the \emph{payoff matrix}, $\mathcal{A} = (a_{ij})$, correspond to the payoff gained by the row agent and fully characterize the game. A subclass of all the possible $2 \times 2$ payoff matrices representing different types of social dilemmas is
\begin{equation}
\label{eq:payoff_matrix}
\begin{array}{ccc}
          & \mbox{ } \!\!\!\!{\rm C} & \!\!\!\!\!\!\! {\rm D} \\
          \begin{array}{c} {\rm C} \\ {\rm D} \end{array} & \!\!\!\!\!\!\left(\!\begin{array}{c} R \\ T
          \end{array}\right. & \left.\!\!\!\!\begin{array}{c} S \\ P
          \end{array}\!\right)\\
          & \mbox{ }  & \mbox{ }%
\end{array}
=
\begin{array}{ccc}
          & \mbox{ } \!\!\!\!{\rm C} & \!\!\!\!\!\!\! {\rm D} \\
          \begin{array}{c} {\rm C} \\ {\rm D} \end{array} & \!\!\!\!\!\!\left(\begin{array}{c} 1 \\ T
          \end{array}\right. & \left.\!\!\!\!\begin{array}{c} S \\ 0
          \end{array}\right)\\
          & \mbox{ }  & \mbox{ }%
\end{array}
= \mathcal{A}
%
\;,
\end{equation}
with $R=1$, $P=0$, $T \in [0,2]$, and $S \in [-1,1]$ \cite{santos-prl-2005,santos-pnas-2006,gintis-book-2009}. For example, when agent $i$ cooperates (\ie{} $s_i = {\rm C}$) and agent $j$ defects ($s_j = {\rm D}$), the former gets payoff $a_{{\rm CD}} = S$, and the latter gets $a_{{\rm DC}} = T$. Matrix $\mathcal{A}$ allows us to study the following games: the Harmony Game (HG), the Stag Hunt (SH), the Hawk and Dove (HD), and the Prisoner's Dilemma (PD) (see Fig.~\ref{fig:games-scheme} for the values of $T$ and $S$ corresponding to each game); albeit in this study we focus on the latter three.
%
%
%
In each round, the total payoff for agent $i$ is given by $\pi(i) = \tfrac{1}{N-1} \sum_{j=1; j\neq i}^N a_{s_i s_j}$. Once all agents have played with all the others, an ordered pair $(i,j)$ of agents is selected uniformly at random to update their strategies. Agent, $i$, with strategy ${\rm X}$ and payoff $\pi_{{\rm X}}$, will adopt the strategy ${\rm Y}$ of agent $j$, with payoff $\pi_{{\rm Y}}$, where ${\rm X,Y} \in \{{\rm C,D}\}$, with a probability $P_{{\rm X} \leftarrow {\rm Y}}$ specified by the so-called \emph{Fermi rule} \cite{blume-gam_eco_beh-1993,szabo-pre-1998}:
\begin{equation}
\label{eq:update_fermi}
P_{{\rm X} \leftarrow {\rm Y}} = \dfrac{1}{1 + e^{-\beta (\pi_{\rm Y} - \pi_{\rm X})}}\,,
\end{equation}
where $\beta \in [0,\infty[$ is an inverse temperature parameter, which is also called the selection pressure. A large value of $\beta$ makes $P_{{\rm X} \leftarrow {\rm Y}}$ more sensitive to the payoff difference, $\pi_{\rm Y} - \pi_{\rm X}$, corresponding to a strong selection pressure. After the strategy updating, the round is completed. Then, all agents reset their payoffs and play the next round. The state of the system is determined by the fraction of cooperative agents, $\fc = N_{\rm C} / N \in [0,1]$, where $N_{\rm C}$ is the number of cooperators. The dynamics continues until the system reaches one of its absorbing states, \ie{} $\fc = \{0,1\}$ \cite{nowak-book-2006,gintis-book-2009}. Committed individuals are modeled as a new class of agents called \emph{zealots}. A zealot always cooperates and, is immune to strategy updating, while other agents may still imitate her by becoming cooperators.

%
%

\section{Results for well-mixed populations}
\label{sec:res_meanfield}

\subsection{Agent-based simulations}
\label{ssec:agent_based_analysis}

\begin{figure}
\centering
\includegraphics[width=0.95\columnwidth]{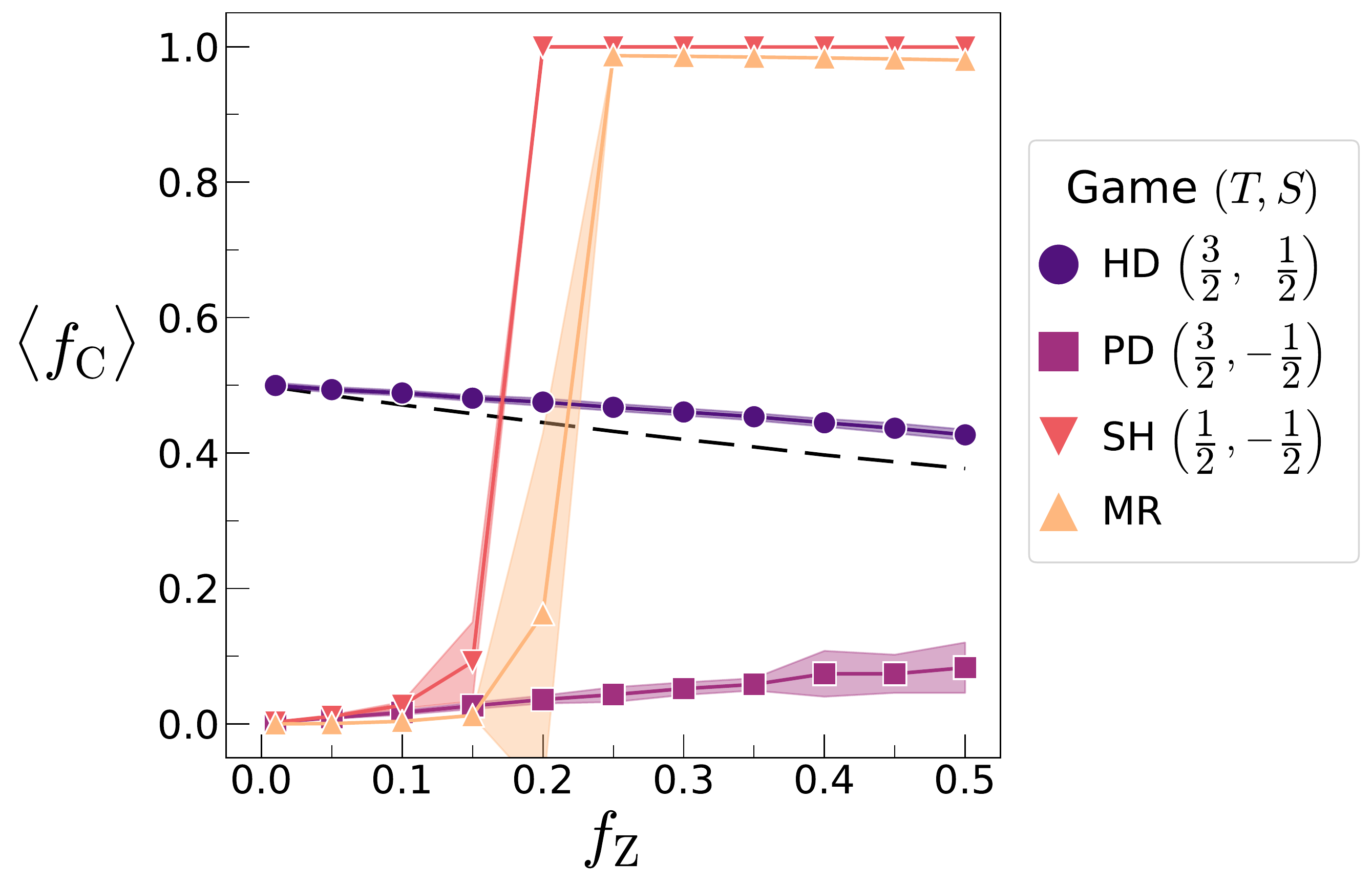}
\caption{Average fraction of cooperation among normal agents in the stationary state, $\avg{\fc}$, against the fraction of zealots, $\fz$. We assume a well-mixed population and set $N = 10^4$ and $\beta = 10$. We also consider the majority rule (MR) model with a neighborhood's size of five. The dashed line represents the analytical result for the HD game in infinite populations (see Appendix~\ref{ssec:infinite_fermi_update} for details). For each $\fz$ value, we average the results over 50 different realizations. Shaded areas represent the standard deviations.}
\label{fig:coop_vs_fz_mf_all_games}
\end{figure}

To examine the extent to which a finite fraction of committed cooperators triggers cooperation, we replace a fraction $\fz = N_{\rm Z} / N \in [0,1/2]$ of normal agents with zealots. Then, we measure the fraction of cooperators among the normal agents, $\fc = N_{\rm C} / \left[(1-\fz) N\right]$, in the stationary state. Figure~\ref{fig:coop_vs_fz_mf_all_games} shows the average fraction of cooperators in the stationary state, denoted by $\avg{\fc}$, for the HD, PD, and SH games for arbitrary choices of $T$ and $S$ (see Appendix~\ref{ssec:ts_exploration} for a comprehensive exploration of the $(T,S)$ space).

%
%
%

%
%
\begin{figure*}[t]
\centering
\includegraphics[width=0.95\textwidth]{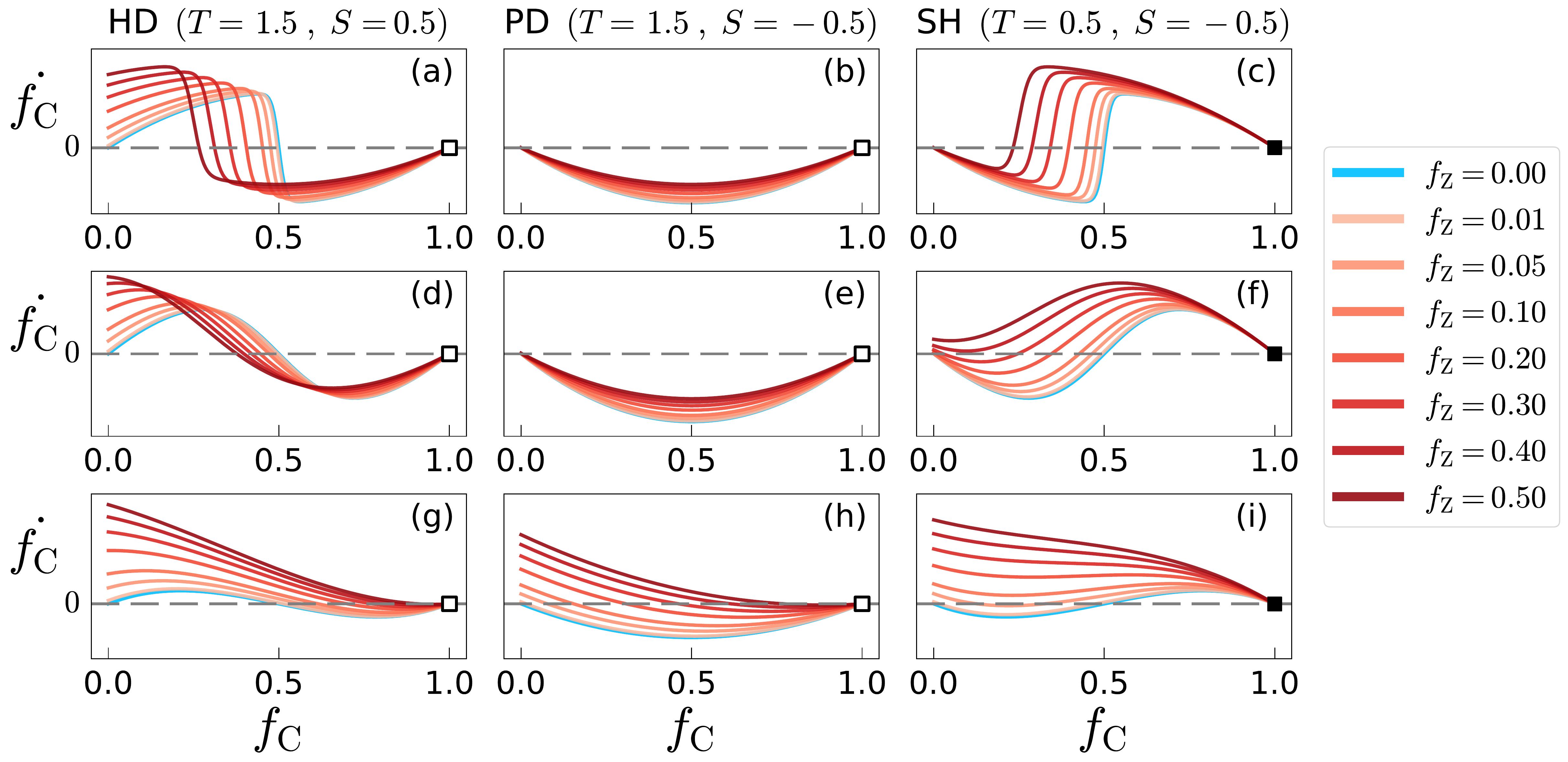}
\caption{Phase portrait of the density of cooperators, $\fc$, undergoing evolutionary dynamics in infinite populations. Each column refers to a different dilemma: HD (panels a, d, and g), PD (b, e, and h), and SH (c, f, and i). We use the Fermi update rule. Each row corresponds to a different selection pressure, namely: $\beta = 100$ (top), $\beta = 10$ (middle), and $\beta = 1$ (bottom).}
\label{fig:results_inf_pop_mf_all_games}
\end{figure*}

Figure~\ref{fig:coop_vs_fz_mf_all_games} indicates that $\fz$ has little effect on $\avg{\fc}$ for the HD and PD games. However, for the SH game we observe at $\fz \approx 0.15$ a sharp transition from a state where cooperation is sustained almost exclusively by zealots (\ie{} $\avg{\fc} \approx 0$), to a fully cooperative state (\ie{} $\avg{\fc} = 1$), denoting the existence of a \emph{critical mass effect} akin to the one observed for the majority rule opinion model \cite{galam-phys_a-2007} and for other opinion dynamics models \cite{mobilia-prl-2003,xie-pre-2011,centola-science-2018}. In contrast, the lack of a critical mass effect observed in the HD and PD games suggests the inability of zealots to initiate a positive feedback mechanism enabling large-scale invasion of cooperators. Furthermore, defectors end up exploiting zealots in the HD game, yielding a reduction in $\fc$ as the fraction of zealots increases \cite{matsuzawa-pre-2016}. This trend is consistent with the theoretical result for infinite well-mixed populations, corresponding to the dashed line in Fig.~\ref{fig:coop_vs_fz_mf_all_games} (see Appendix~\ref{ssec:infinite_fermi_update} for the details).

\subsection{Infinite-size populations}
\label{ssec:inf_pop_analytical_analysis}

\subsubsection{Methods and overall results}
\label{sssec:inf_pop_meth_and_overall}

To address why zealots trigger the onset of cooperation in the SH but not in the HD and PD games, we study the evolutionary dynamics in terms of the concentration of cooperators, $\fc$, in the thermodynamic limit. The following critical mass results are qualitatively the same for finite populations (Fig.~\ref{fig:mf_finite_size_fermi_allb}), as well as for the so-called birth-death rule, which is equivalent to replicator dynamics in infinite populations \cite{nowak-book-2006} (see Fig.~\ref{fig:mf_replicator_all_w} and Appendix~\ref{ssec:infinite_replicator_update} for details). Given an infinite population, the evolutionary dynamics are described by
\begin{equation}
\label{eq:evol_cooperation}
\dot{\fc} = \mathcal{P}_{{\rm D} \rightarrow {\rm C}} - \mathcal{P}_{{\rm C} \rightarrow {\rm D}} \,,
\end{equation}
where $\dot{\fc}$ denotes $\fc$'s derivative with respect to time. Quantity $\mathcal{P}_{{\rm X} \rightarrow {\rm Y}}$ is the probability per unit time that a pair of agents having strategies ${\rm X}$ and ${\rm Y}$ is selected for updating, and then agent with strategy ${\rm X}$ changes it to ${\rm Y}$. We have assumed that, the change of the strategy obeys to the Fermi rule given by Eq.~\eqref{eq:update_fermi}. Therefore, one obtains
\begin{equation}
\label{eq:dyn_inf_pop_fermi}
\dot{\fc} = \dfrac{(1 - \fc)}{(1 + \fz) \bigl(1+e^{\beta\alpha}\bigr)} \Bigl[ (\fc + \fz ) e^{\beta\alpha} - \fc \Bigr] \,,
\end{equation}
where
\begin{equation}
\label{eq:payoff_diff}
\alpha = \pic - \pid = \dfrac{1}{1 + \fz} \Bigl[ (\fc + \fz) (1-T) + (1 - \fc)S \Bigr] \,,
\end{equation}
is the difference between the payoffs of a cooperator and a defector.

We search then for the equilibria of the dynamics described by Eq.~\eqref{eq:dyn_inf_pop_fermi}, \ie{} ${\fc}^\star$ such that $\bigl.\dot{\fc}\bigr\vert_{\fc = {\fc}^\star} \!\! = 0$. This is effectively done in Fig.~\ref{fig:results_inf_pop_mf_all_games}, where we plot $\dot{\fc}$ as a function of $\fc$ (\ie{} we draw the phase portrait) for the three types of games and three selection pressure values, $\beta = \{100, 10, 1\}$. The hue of the lines corresponds to different values of $\fz$ including the case of zealots' absence (blue line). The solid square represents the fully cooperative absorbing state ($\fc = 1$), and the other equilibria correspond to the intersection(s) of the curves with $\dot{\fc} = 0$ (dashed lines). We first observe that ${\fc}^\star = 0$ (\ie{} absence of cooperation) exists if and only if zealots are absent. Therefore, zealots induce a positive fraction, which may be small, of cooperators among normal agents (see Appendix~\ref{ssec:infinite_fermi_update} for the proof). Second, equilibrium ${\fc}^\star = 1$ (\ie{} fully cooperative state) exists for any $\fz$ and is either unstable (HD and PD games) or stable (SH game). Third, there is at most one intermediate solution, $\gamma =\fz \exp(\beta \alpha) / [1-\exp(\beta \alpha)]$, whose position depends on the dilemma's type, $\fz$, and the selection pressure, $\beta$. In the following, we discuss the role of the selection pressure $\beta$ on the critical mass effect (if it exists) for each game.


%
%

\subsubsection{HD game}
\label{sssec:mf_inf_pop_hd}

In the HD game, for a strong selection pressure (Fig.~\ref{fig:results_inf_pop_mf_all_games}(a) and (d)), ${\fc}^\star = \gamma$ is always stable, but its position moves towards $\fc = 0$ as $\fz$ becomes larger (in agreement with the results shown in Fig.~\ref{fig:coop_vs_fz_mf_all_games}). This displacement is due to the fact that defectors exploit zealots thus helping defection to thrive. In this sense, cooperative zealotry has a detrimental effect on cooperation as the more zealots there are in the population, the less cooperation we observe. Moreover, a stronger selection pressure (Fig.~\ref{fig:results_inf_pop_mf_all_games}(a)) exacerbates the exploitation of cooperators compared to a milder selection pressure (Fig.~\ref{fig:results_inf_pop_mf_all_games}(d)). 

For a weak selection pressure, the results are opposite to those for stronger selection in the sense that ${\fc}^\star = \gamma$ moves towards $\fc = 1$ (as opposed to $\fc = 0$) as one adds zealots (Fig.~\ref{fig:results_inf_pop_mf_all_games}(g)). This means that reducing the selection pressure reduces the effects of payoff difference, $\alpha$, making cooperation more appealing and boosting the ability of zealots to trigger the emergence of cooperation.

%
%

\subsubsection{PD game}
\label{sssec:mf_inf_pop_pd}

In the PD game, the position of ${\fc}^\star = \gamma$ is slightly positive for strong levels of selection (Figs.~\ref{fig:results_inf_pop_mf_all_games}(b) and (e)), such that zealots induce some, albeit small, cooperation in agreement with the observation made in Refs.~\cite{masuda-scirep-2012,mobilia-pre-2012}. For a weak selection pressure, the equilibrium ${\fc}^\star = \gamma$ exists for any $\fz > 0$ and is stable (Fig.~\ref{fig:results_inf_pop_mf_all_games}(h)). A reduced selection pressure makes the payoff difference less relevant, and ${\fc}^\star = \gamma$ moves towards $\fc = 1$ as $\fz$ increases. Nevertheless, we do not observe a critical mass effect.

%
%

\subsubsection{SH game}
\label{sssec:mf_inf_pop_sh}

In the SH game (Figs.~\ref{fig:results_inf_pop_mf_all_games}(c), (f), and (i)), ${\fc}^\star = \gamma$ is unstable when it exists, and the response to zealots is richer than in the HD and PD games. Under a strong selection pressure, we do not observe a critical mass effect. The dynamics is bistable meaning that, depending on the initial condition, it attains one of the two stable equilibria in which either ${\rm C}$ or ${\rm D}$ is the majority (Fig.~\ref{fig:results_inf_pop_mf_all_games}(c)). This dynamics is qualitatively the same as that for $\fz = 0$, in which case the stable equilibria are located at ${\fc}^\star = 0$ and ${\fc}^\star = 1$, and the unstable equilibrium is located at ${\fc}^\star = \gamma = 1/2$. However, as $\fz$ increases, ${\fc}^\star = \gamma$ moves towards the left indicating that zealots make cooperation more appealing, favoring the convergence to ${\fc}^\star =1$. For a mild selection pressure, $\beta = 10$, for $\fz < {\fz}^\star \approx 0.4$ the dynamics remains bistable (Fig.~\ref{fig:results_inf_pop_mf_all_games}(f)). This result is similar to that for $\beta = 100$. However, for $\fz \geq {\fz}^\star$ the system undergoes a saddle-node bifurcation, through which the left stable equilibrium and ${\fc}^\star = \gamma$ coalesce and disappear, leaving ${\fc}^\star = 1$ as the only equilibrium at $\fz > {\fz}^\star$. The saddle-node bifurcation confirms the existence of a critical mass effect. We report an extensive exploration of the $(T,S)$ space in Fig.~\ref{fig:mf_numeric_coop} of Appendix~\ref{ssec:ts_exploration}. Finally, a weak selection pressure, $\beta = 1$, reduces the critical mass threshold, ${\fz}^\star$, from $\approx 0.4$ to $\approx 0.08$ (Fig.~\ref{fig:results_inf_pop_mf_all_games}(i)).

\subsubsection{Summary}
\label{sssec:mf_inf_pop_summary}

In a nutshell, reducing the selection pressure, $\beta$, does not spawn a critical mass effect in the HD and PD games. However, doing so still fosters higher levels of cooperation in all the three types of games, and reduces the threshold for the critical mass effect in terms of $\fz$ in the SH game. Moreover, reducing $\beta$ attenuates the role of payoff difference, $\alpha$, on the fixation of cooperation, triggering the shift of the intermediate equilibrium, $\gamma$. Although zealots do not trigger the transition to full cooperation in the HD and PD games, they elongate the transient time needed to reach the equilibrium (Fig.~\ref{fig:mf_numeric_conv_time} in Appendix~\ref{ssec:ts_exploration} and Ref.~\cite{nakajima-math_bio-2015}). We also underline that, apart from the selection pressure and game type, the critical mass effect depends on the type of update rule. Under replicator dynamics, in fact, the critical mass effect with ${\fz}^\star = 0.3$ exists regardless of whether the selection pressure is strong or weak, and under all types of games (Fig.~\ref{fig:mf_replicator_all_w} in Appendix~\ref{ssec:infinite_replicator_update}).

%
%

\section{Results in network-based populations}
\label{sec:res_networks}

%
%
\begin{figure*}[t]
\centering
\includegraphics[width=0.95\textwidth]{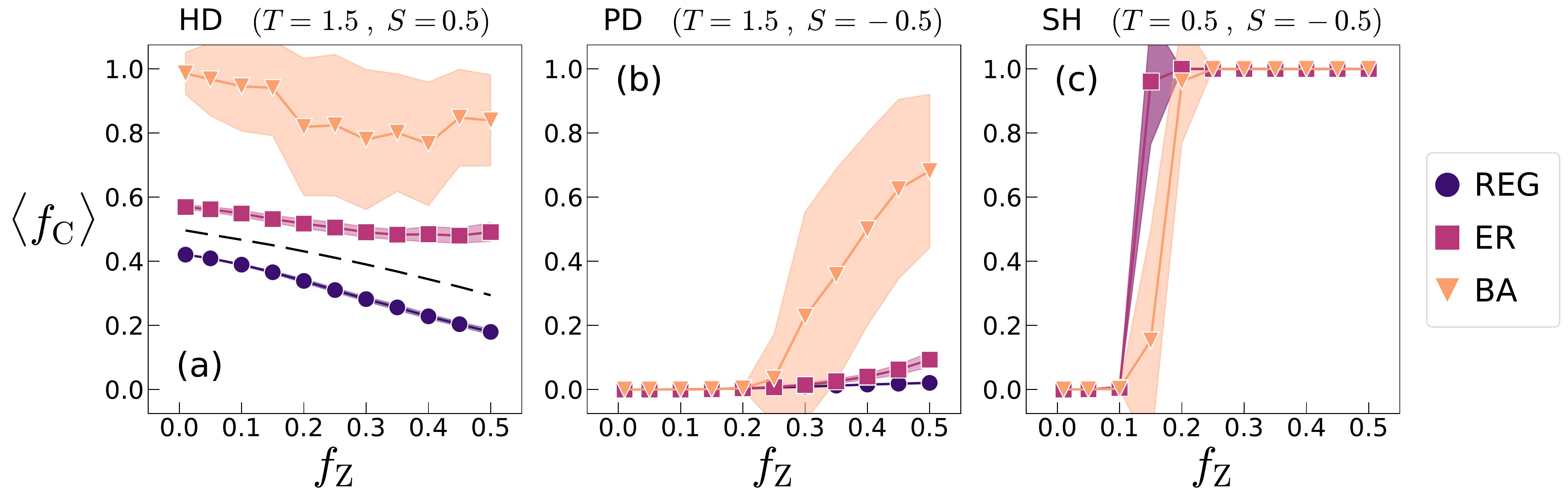}
\caption{Effect of zealotry on the emergence of cooperation on networks. We plot the average fraction of cooperators among normal agents, $\avg{\fc}$, against the fraction of zealots, $\fz$ for (a) HD, (b) PD, and (c) SH. We used a Fermi update rule with $\beta = 10$. We consider three types of networks, namely, random regular lattice (REG), Erd\H{o}s-Reny\'i (ER), and Barab\'asi-Albert networks (BA). We set $N=1000$, and $\avg{k} = 6$. In the BA networks, the size of the initial seed $m_0 = 3$, and number of edges added per step $m = 3$. The dashed line represents the analytical result for the HD game in finite populations (see Appendix~\ref{sec:evol_coop_finite_size}). Results are averaged over 50 different realizations.}
\label{fig:results_all_games_networks}
\end{figure*}

It has been shown that the structure of the interactions plays a pivotal role in the emergence of cooperation \cite{nowak-science-2006,szabo-phys_rep-2007}. For this reason, we consider populations of agents where the interactions among them are encoded into a network. One of the main results of evolutionary games played on networks is that the heterogeneity of the number of neighbors per agent (\ie{} the degree) favors the emergence of cooperation \cite{santos-prl-2005,santos-pnas-2006}. To assess whether the heterogeneity in the degree plays a role also in the ability of zealots to trigger a critical mass effect, we consider networks with an increasing level of degree heterogeneity, \ie{} the random regular lattice (REG), the Erd\H{o}s-Reny\'i (ER), and B\'arabasi-Albert (BA) networks \cite{erdos-paper-1959,barabasi-science-1999,latora-book-2017}. In REG networks all agents have exactly the same degree. Under this constraint, the edges are uniformly randomly wired. In ER networks, the edges are uniformly randomly wired under the condition that the degree distribution is a Poisson distribution. Equivalently, each pair of nodes is adjacent to each other by an edge with a fixed probability, independently of the other node pairs. Finally, in BA networks nodes are added one at a time, and connections are established following the so-called preferential attachment rule. In BA networks, the degree is heterogeneously distributed and spans several orders of magnitude: the degree distribution is a power-law function. All networks have the same number of nodes, $N=1000$, and average degree, $\avg{k}=6$.

In Fig.~\ref{fig:results_all_games_networks}, we report the fraction of cooperators among normal agents in the three types of games for REG, ER, and BA networks. Agents' payoffs are computed according to the so-called additive payoff scheme, \ie{} the payoff of an agent $i$ is $\pi(i) = \sum_{j \in \mathcal{N}_i} a_{s_i s_j}$, where $\mathcal{N}_i$ is the set of neighbors of $i$ \cite{santos-prl-2005}. We have confirmed that the results remain qualitatively the same for REG and ER also when we normalize $\pi(i)$ by agent $i$'s degree (\ie{} we use the average payoff scheme), although this is not the case of BA networks (see Appendix~\ref{sec:networks-numerics} for details). We consider a fraction $\fz$ of uniformly randomly selected nodes to be zealots, whereas the other nodes are initialized as defectors.

In the HD game, BA networks, but not the REG and ER networks, attain almost full cooperation for $\fz = 0$, which is in agreement with previous results \cite{abramson-pre-2001,santos-pnas-2006,roca-phys_lif_rev-2009}. An increase in $\fz$ allows defectors to exploit normal cooperators regardless of the networks' structure (Fig.~\ref{fig:results_all_games_networks}(a)), in agreement with results for well-mixed populations (Fig.~\ref{fig:coop_vs_fz_mf_all_games}). In the PD game, BA networks display a clear critical mass effect, with a sensible increase in cooperation for ${\fz}^\star > 0.25$. We set $T = 1.5$ and $S = −0.5$ such that cooperation would be absent in the BA networks without zealots. We find that the presence of zealots considerably enhances cooperation in the PD game (Fig.~\ref{fig:results_all_games_networks}(b)). Such enhancement of cooperation confirms that degree heterogeneity plays a crucial role in how zealots trigger the emergence of cooperation. Finally, in the SH game (Fig.~\ref{fig:results_all_games_networks}(c)), the different structure of the network slightly anticipates the transition to $\fc = 1$ in terms of $\fz$, as compared to the case of well-mixed populations. Our extensive exploration of the $(T, S)$ space supports the generality of these results (Figs.~\ref{fig:all_nets_coop_nodegq} and \ref{fig:all_nets_coop_degq}).

%
%

\section{Conclusions}
\label{sec:conclusions}

In this paper, we have studied how committed individuals induce qualitative and quantitative changes in evolutionary dynamics under different types of social dilemma. We have considered three evolutionary dilemmas, and observed that only the SH game displays a clear critical mass effect, whereas in the other two dilemmas we need to reduce the selection pressure, change the update rule, or consider heterogeneous networks to observe a critical mass effect. The presence of a critical mass in terms of the fraction of zealots in the SH game is in line with the observations made for other coordination dynamics such as the majority rule model \cite{galam-phys_a-2007}, the naming game \cite{xie-pre-2011,mistry-pre-2015,waagen-pre-2015,centola-science-2018}, and the voter model \cite{mobilia-prl-2003}.

Given our analytical underpinning, an experimental validation of our findings involving populations of human subjects \cite{grujic-pone-2010} -- akin to the one done by Centola \etal for the naming game \cite{centola-science-2018} -- would be desirable. On the other hand, the extensions to evolutionary vaccination scenarios \cite{liu-pre-2012} where zealots contribute in part to herd immunization and elicit defection, \ie{} not taking immunization, among normal agents (in agreement with the phenomenology observed in the HD game) would be straightforward. In the evolutionary synchronization dynamics based on the so-called evolutionary Kuramoto dilemma \cite{antonioni-prl-2017} (which is akin to a coordination dynamics), the existence of a critical mass -- in analogy with the SH game -- is expected to enable the emergence of either global or chimera synchronized states. Such states would not be observed in the absence of zealots. Hence, the addition of zealots to populations of evolutionary oscillators could be seen as the evolutionary analog of pinning in control theory \cite{delellis-ieee-2010}. Our findings may also be useful for understanding why zealots succeed in suppressing oscillations in the rock-scissor-paper game \cite{szolnoki-pre-2016}, whether or not a non-negligible fraction of zealots is required for the emergence of collective coordination in human \cite{dyer-animal_beh-2008} and animal \cite{couzin-science-2011} behavior, cooperative transport in ants \cite{feinerman-nphys-2018,mc_creery-insec_soc-2017}, and cell migration \cite{kabla-j_roy_soc_int-2012}. For the latter two dynamics, an elongated correlation length plays a role on coordination \cite{kabla-j_roy_soc_int-2012,feinerman-nphys-2018} similar to a low selection pressure in our framework. Further developments of the present work include studies of how tiny fractions of zealots may affect the so-called fixation probability (\ie{} the probability that the system ends up in the fully cooperative absorbing state) and the fixation time (\ie{} the time needed to reach the adsorbing state). Finally, we have considered only a random placement of zealots on the nodes of the network. However, placing zealots on nodes ranked according to their topological properties can reduce the critical fraction of zealots needed to attain full consensus \cite{arendt-comp_math_org_theo-2015}.
%


%
%
\begin{acknowledgments}
The authors thank A. Baronchelli for his talk which inspired this work. A.C. acknowledges the support of the Spanish Ministerio de Ciencia e Innovacion (MICINN) through Grant IJCI-2017-34300. A.C. and N.M. acknowledge the support of Cookpad Limited. This work was carried out using the computational facilities of the Advanced Computing Research Centre, University of Bristol - \url{http://www.bristol.ac.uk/acrc/}. Graphics have been prepared using the Matplotlib Python package \cite{hunter-matplotlib-2007}.
\end{acknowledgments}

%
%

\appendix

%
%


\section{Evolutionary dynamics in finite populations}
\label{sec:evol_coop_finite_size}

In finite populations, we can study the onset of cooperation under the presence of zealots using one-dimensional random walks \cite{nowak-book-2006,antal-bull_math_bio-2006,nakajima-math_bio-2015}. Let us consider a population of size $N = N^\prime + z$, where $N^\prime$ is the number of normal agents, and $z = \fz \, N$ is the number of zealous agents. Our goal is to study the evolution of the number of cooperators among normal agents, $N_{\rm C} \equiv i \; (0 \leq i \leq N^\prime)$, given a fraction of zealots, $\fz$. For the payoff matrix given by Eq.~\eqref{eq:payoff_matrix}, the payoffs associated with cooperation and defection are equal to
\begin{align}
\label{eq:payoffs_finite_pop_pc}
\pic &= \dfrac{\left( i + z - 1 \right) + \left( N^\prime - i \right)S}{N - 1}\\
\intertext{and}
\label{eq:payoffs_finite_pop_pd}
\pid &= \dfrac{\left( i + z\right)T}{N - 1}\,,
\end{align}
%
%
%
%
\begin{figure}[t!]
\centering
\includegraphics[width=0.7\columnwidth]{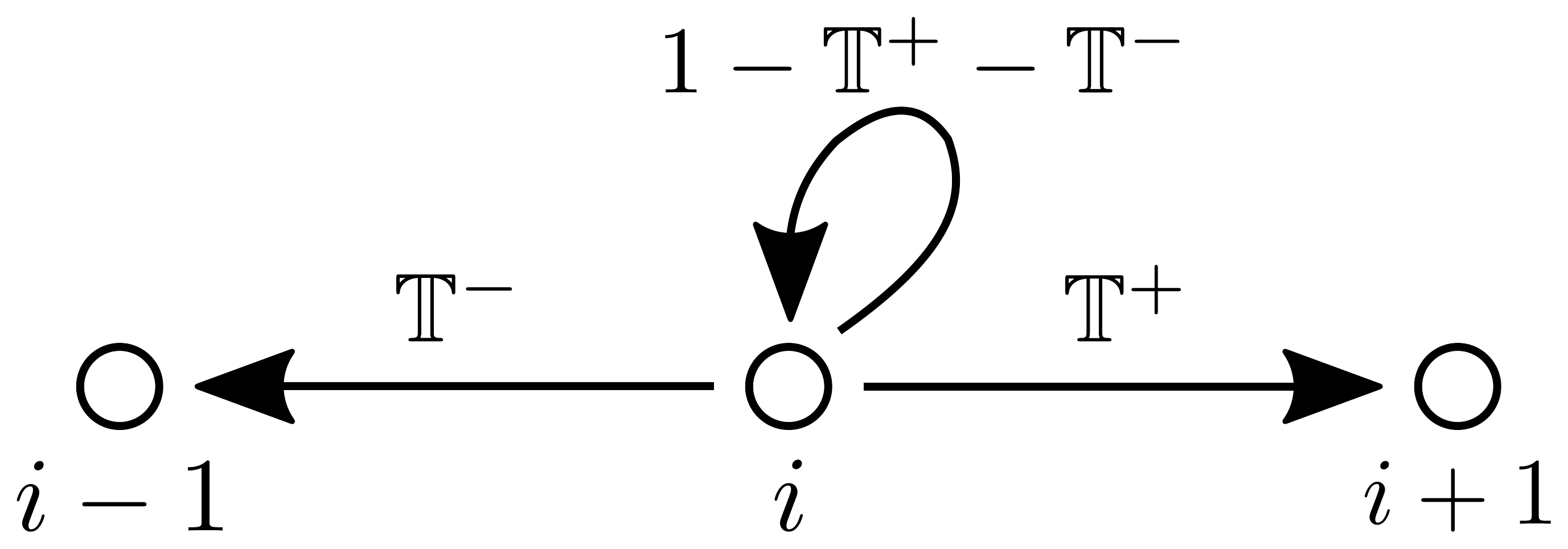}
\caption{Schematic representation of the possible transitions that a finite population with $i$ cooperators can undergo in one step under evolutionary dynamics.}
\label{fig:markov-process}
\end{figure}
\noindent respectively. The evolutionary dynamics is equivalent to a one-dimensional random walk on a finite line whose position is identified with the number of normal cooperators, $0 \leq i \leq N^\prime$. Figure \ref{fig:markov-process} shows the possible state transitions in each strategy updating, where $\mathbb{T}^-$, $\mathbb{T}^+$, and $1 - \mathbb{T}^+ - \mathbb{T}^-$ are the probabilities that $i$ decreases by one, increases by one, and remains the same after a single strategy updating, respectively. The probability that $i$ increases (decreases) by one is equal to the product of the probability of choosing a pair of agents with different strategies, $P_{\text{ch}}$, and the probability that one of the agents updates its strategy, $\mathcal{P}_{{\rm D} \leftarrow {\rm C}}$ ($\mathcal{P}_{{\rm C} \leftarrow {\rm D}}$). By keeping in mind that zealots do not update their strategy but can induce defectors to become cooperators, under the Fermi rule one obtains
\begin{align}
\label{eq:ladder_operators_finite_size_tp}
\mathbb{T}^+ &=  P_{\text{ch}} \cdot \mathcal{P}_{{\rm D} \leftarrow {\rm C}} = \dfrac{\left( i + z \right) \left(N^\prime - i \right)}{N^2} \dfrac{1}{1 + e^{-\beta (\pic - \pid)}} \,, \\
\label{eq:ladder_operators_finite_size_tm}
\mathbb{T}^- &=  P_{\text{ch}} \cdot \mathcal{P}_{{\rm C} \leftarrow {\rm D}} = \dfrac{i \left(N^\prime - i \right)}{N^2} \dfrac{1}{1 + e^{-\beta (\pid - \pic)}} \,,
\end{align}
where we have used $N (N-1) \simeq N^2$ assuming a large $N$. Then, the bias in the random walk reads
\begin{multline}
\label{eq:begin_random_walk_bias_finite_pop}
\mathbb{T}^+ - \mathbb{T}^- = \dfrac{\left( i + z \right) \left(N^\prime - i \right)}{N^2} \dfrac{1}{1 + e^{-\beta (\pic - \pid)}}\\ - \dfrac{i \left(N^\prime - i \right)}{N^2} \dfrac{1}{1 + e^{-\beta (\pid - \pic)}}\,.
\end{multline}
It is convenient to compute the payoff difference\newline $\alpha = \pic - \pid$ using Eqs.~\eqref{eq:payoffs_finite_pop_pc} and \eqref{eq:payoffs_finite_pop_pd}, yielding
\begin{equation}
\label{eq:payoff_diff_finite_size}
\alpha = \pic - \pid = \dfrac{\left( 1 - S - T \right) i + \left( 1 - T \right)z + \left( N^\prime S - 1\right)}{N - 1}\,.
\end{equation}
In terms of $\alpha$, Eq.~\eqref{eq:begin_random_walk_bias_finite_pop} is rewritten as
\begin{equation}
\label{eq:random_walk_bias_finite_pop}
\mathbb{T}^+ - \mathbb{T}^- = \dfrac{\left(N^\prime - i \right) \left[ \left( i + z \right)  e^{\beta\alpha} - i \right]} {N^2 \left[ 1 + e^{\beta\alpha} \right]} \,.
\end{equation}
The equilibria of the evolutionary dynamics are the values of $i^\star$ such that $\mathbb{T}^+ - \mathbb{T}^- = 0$. Given Eq.~\eqref{eq:random_walk_bias_finite_pop}, finding the equilibria is equivalent to finding the solutions of
\begin{equation}
\left(N^\prime - i \right) \left[ \left( i + z \right)  e^{\beta\alpha} - i \right] = 0 \,.
\end{equation}
%
%
%
%
%
%
%
\begin{figure*}[t!]
\centering
\includegraphics[width=0.95\textwidth]{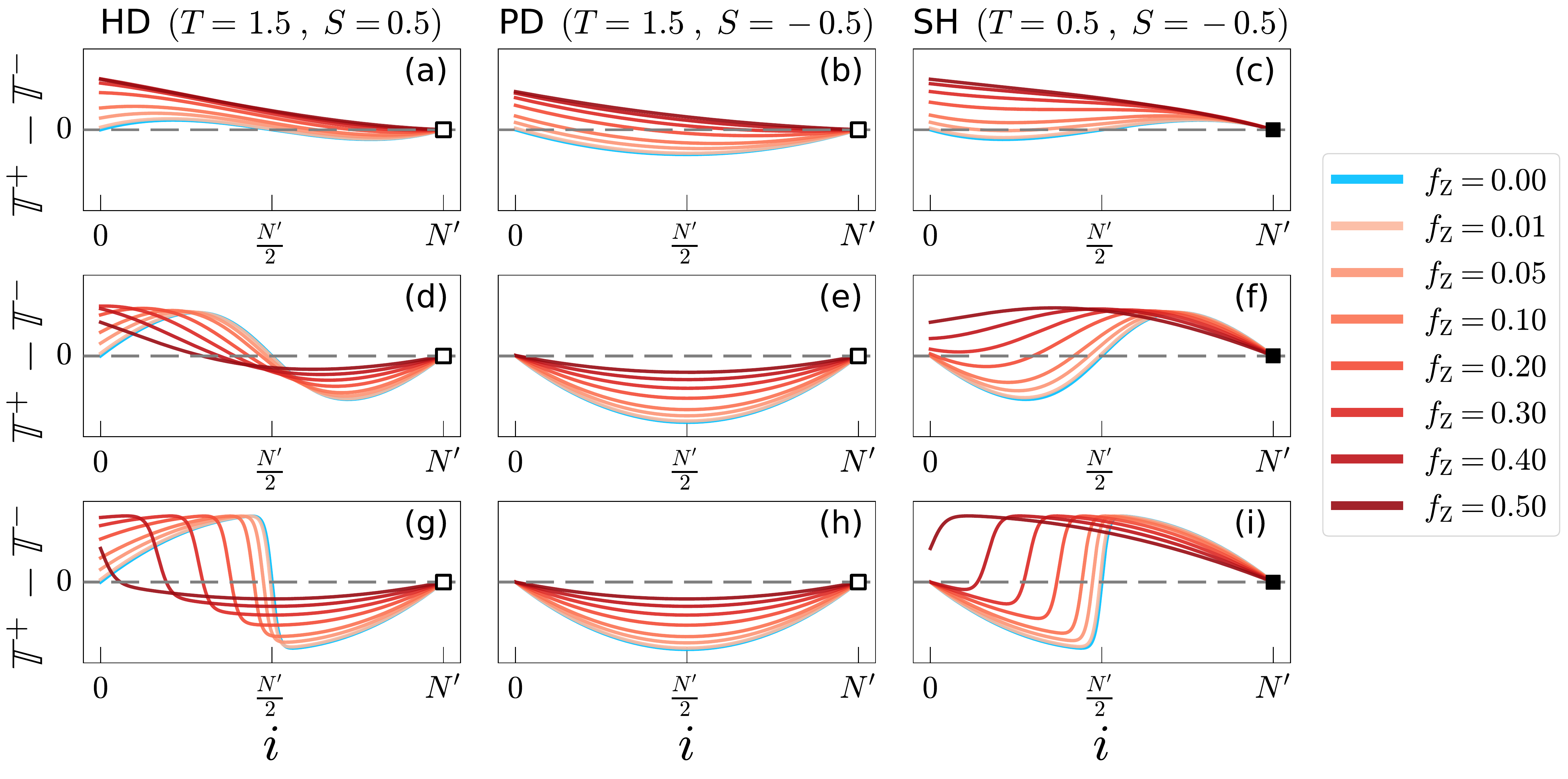}
\caption{Random walk's bias, $\mathbb{T}^+ - \mathbb{T}^-$ (Eq.~\eqref{eq:random_walk_bias_finite_pop}), as a function of the number of cooperators among normal agents, $i$. Each column accounts for a different dilemma, namely, HD (panels a, d, and g), PD (b, e, and h), and SH (c, f, and i). The hue of the solid lines corresponds to the fraction of zealots, $\fz$, in the population, with the case without zealots being shown in cyan. The square at $i = N^\prime$ represents the fully cooperative absorbing equilibrium, and the filling color denotes whether the equilibrium is stable (black) or unstable (white). The results are for the Fermi update rule. Each row corresponds to a different value of $\beta$, namely, $\beta = 1$ (top row), $\beta = 10$ (middle row), and $\beta = 100$ (bottom row). The population size is $N = 10^5$.}
\label{fig:mf_finite_size_fermi_allb}
\end{figure*}
%
%
Therefore, $i^\star = N^\prime$, corresponding to the fully cooperative state, is an equilibrium. To find other equilibria, we start by rewriting the payoff difference, $\alpha$, in terms of $i$ as follows
\begin{align}
\alpha &= \dfrac{\left( 1 - S - T \right) i + \left( 1 - T \right)z + \left( N^\prime S - 1\right)}{N - 1} \nonumber \\
%
%
       \label{eq:def_alpha_lambda}
       &= \dfrac{\lambda_1 + \lambda_2 \, i}{\beta} \,,
\intertext{where}
\label{eq:def_lambda1}
\lambda_1 &= \beta \, \dfrac{\left( 1 - T \right)z + \left( N^\prime S - 1\right)}{N - 1} \\
\intertext{and}
\label{eq:def_lambda2}
\lambda_2 &= \beta \, \dfrac{1 - S - T}{N - 1}\,.
\end{align}
In terms of $\lambda_1$ and $\lambda_2$, we obtain
\begin{equation}
\label{eq:stationary_state_finite_pop}
\left( i + z \right)  e^{\beta\alpha} - i = i \, \left( e^{\lambda_1} e^{\lambda_2 \, i} -1 \right) + z \, e^{\lambda_1} e^{\lambda_2 \, i} = 0 \,.
\end{equation}
Therefore, $i^\star = 0$ is solution of Eq.~\eqref{eq:stationary_state_finite_pop} if and only if $z = 0$. This means that zealots always induce some cooperation among normal agents. Furthermore, in agreement with the behavior observed in absence of zealots for SH and HD games, we expect to observe another equilibrium $0 < i^\star < N^\prime$. However, finding a closed-form expression for such an equilibrium (\ie{} the solution of Eq.~\eqref{eq:stationary_state_finite_pop}) is not easy because Eq.~\eqref{eq:stationary_state_finite_pop} is a transcendental equation. After identifying all the equilibria, we examine their stability. Equilibrium $i^\star = 1$ is stable and corresponds to full cooperation. Let us analyze the stability of other equilibria by focusing on the sign of the random-walk bias given by Eq.~\eqref{eq:random_walk_bias_finite_pop}. The sign exclusively depends on the numerator's sign because the denominator of Eq.~\eqref{eq:random_walk_bias_finite_pop} is always positive. Specifically, we examine whether
\begin{equation}
\label{eq:rw_bias_sign}
\underbracket{\left(N^\prime - i \right)}_{\geq 0} \Bigl[ \; \overbracket{\underbracket{\left( i + z \right)}_{\geq 0}  \underbracket{e^{\beta\alpha}}_{\geq 0} - \underbracket{\, i \,}_{\geq 0}}^{\Lambda} \; \Bigr] \lessgtr 0 \,.
\end{equation}
Let us concentrate on the case $\Lambda > 0$, \ie
\begin{gather}
\label{eq:inequality_lambda_gt0_1}
\left( i + z \right) e^{\beta\alpha} > i \;.
\intertext{By substituting Eq.~\eqref{eq:def_alpha_lambda} into Eq.~\eqref{eq:inequality_lambda_gt0_1} and taking the logarithms of both sides, one obtains}
\label{eq:inequality_lambda_gt0_2}
\lambda_1 + \lambda_2 \, i > \ln \left( \dfrac{i}{i+z} \right)\,,\\
\intertext{By substituting Eqs.~\eqref{eq:def_lambda1} and \eqref{eq:def_lambda2} into Eq.~\eqref{eq:inequality_lambda_gt0_2}, one obtains}
\label{eq:cond_positive_bias_finite_pop}
\beta \dfrac{1 - S - T}{N^\prime + z - 1} \, i + \beta \left[ \dfrac{\left( 1 - T\right) \,z + \left(N^\prime S - 1 \right)}{N^\prime + z - 1} \right] > \ln \left( \dfrac{i}{i+z} \right) \,.
\end{gather}
Similarly, $\Lambda < 0$ leads to
\begin{equation}
\label{eq:cond_negative_bias_finite_pop}
\beta \dfrac{1 - S - T}{N^\prime + z - 1} \, i + \beta \left[ \dfrac{\left( 1 - T\right) \,z + \left(N^\prime S - 1 \right)}{N^\prime + z - 1} \right] < \ln \left( \dfrac{i}{i+z} \right) \,.
\end{equation}
For $i = 0$, inequality \eqref{eq:cond_positive_bias_finite_pop} always holds true. This confirms that $i^\star = 0$ is a reflective equilibrium and that zealots always generate some cooperation.

In Fig.~\ref{fig:mf_finite_size_fermi_allb} we report $\mathbb{T}^+ - \mathbb{T}^-$ given by Eq.~\eqref{eq:random_walk_bias_finite_pop} as a function of the number of cooperators among normal agents, $i$, for three values of $\beta$, three types of game (\ie{} HD, PD, and SH), and $N = 10^5$. Figure \ref{fig:mf_finite_size_fermi_allb} suggests that the results are qualitatively the same as those for the infinite well-mixed populations reported in Figs.~\ref{fig:results_inf_pop_mf_all_games}.


\section{Evolutionary dynamics in infinite populations}
\label{sec:evol_coop_infinite_size}

In this section, we analyze infinite populations following the approach introduced in Ref.~\cite{masuda-scirep-2012}. We study the dynamics of the fraction of cooperators among normal agents, $\dot{\fc}$, where the dot denotes the time derivative, as a function of $\fc$ (\ie{} we plot the phase portrait \cite{strogatz-book-1994}), when we introduce a fraction, $\fz \geq 0$, of cooperative zealots into the population. For strategy updating, we consider the Fermi rule \cite{szabo-pre-1998}, and the replicator dynamics \cite{nowak-book-2006}.

\subsection{Fermi rule}
\label{ssec:infinite_fermi_update}

The average payoff of a cooperator, $\pic$, and that of a defector, $\pid$, are given by
\begin{align}
\label{eq:payoff_coop_def_fermi}
\pic &= \dfrac{\Bigl[ \left( \fc + \fz \right) + \left( 1 - \fc \right) S \Bigr]}{1+\fz}
\intertext{and}
\pid &= \dfrac{\left( \fc + \fz \right) T }{1+\fz} \,,
\end{align}
respectively. Thus
\begin{equation}
\label{eq:var_fc_dot_inf_pop}
\dot{\fc} = \mathbb{T}^+ - \mathbb{T}^- = \mathcal{P}_{{\rm D} \rightarrow {\rm C}}  - \mathcal{P}_{{\rm C} \rightarrow {\rm D}} \;,
\end{equation}
where $\mathcal{P}_{{\rm X} \rightarrow {\rm Y}}$ is the joint probability that one selects a pair of agents with strategies $({\rm X,Y})$ uniformly at random and that the agent with strategy ${\rm X}$ adopts strategy ${\rm Y}$ under the Fermi rule. These probabilities read
\begin{align}
\label{eq:prob_update_fermi_1}
\mathcal{P}_{{\rm D} \rightarrow {\rm C}} &= \dfrac{\left( \fc + \fz \right)\left( 1 - \fc \right)}{1+\fz} \, \dfrac{1}{1 + e^{-\beta\alpha}}\,, \\
\label{eq:prob_update_fermi_2}
\mathcal{P}_{{\rm C} \rightarrow {\rm D}} &= \dfrac{\fc \left( 1 - \fc \right)}{1+\fz} \, \dfrac{1}{1 + e^{\beta\alpha}}\,,
\end{align}
where
\begin{equation}
\label{eq:payoff_diff_fermi}
\alpha \equiv \pic - \pid = \dfrac{1}{1+\fz} \Bigl[ \left( 1 - T \right) \left( \fc + \fz \right) + \left( 1 - \fc \right) S \Bigr] \,.
\end{equation}
By substituting Eqs.~\eqref{eq:prob_update_fermi_1} and \eqref{eq:prob_update_fermi_2} into Eq.~\eqref{eq:var_fc_dot_inf_pop}, one obtains
\begin{equation}
\label{eq:evol_coop_fermi}
\dot{\fc} = \dfrac{1-\fc}{\left(1+\fz\right) \left( 1 + e^{\beta\alpha} \right)} \left[ \left( \fc + \fz \right)e^{\beta\alpha} - \fc \right] \,.
\end{equation}
At equilibrium, ${\fc}^\star$ (\ie{} $\dot{\fc} = 0$), one obtains
\begin{equation}
\label{eq:equilibria_fermi}
{\fc}^\star = 1  \quad\qquad\text{or}\quad\qquad  {\fc}^\star = \dfrac{\fz \, e^{\beta\alpha}}{1 - e^{\beta\alpha}} \,.
\end{equation}
In the absence of zealots (\ie{} $\fz = 0$), we have two equilibria ${\fc}^\star = 0$ and ${\fc}^\star = 1$. When $\fz > 0$, ${\fc}^\star = 1$ is still an equilibrium but ${\fc}^\star = 0$ is not, and it is replaced by a new equilibrium located at $0 < {\fc}^\star < 1$. The behavior of $\dot{\fc}$ against $\fc$ for different values of $\beta$ and three types of dilemmas is displayed in Fig.~\ref{fig:results_inf_pop_mf_all_games}.


\subsection{Replicator dynamics}
\label{ssec:infinite_replicator_update}

The replicator dynamics constitutes the infinite size counterpart of the Moran rule \cite{moran-book-1962,nowak-book-2006}. Under the Moran process, we first select an agent called the \emph{child} uniformly at random among the normal agents. Second, we choose an agent called the \emph{parent} randomly with the probability proportional to the agent's payoff. Third, the parent's strategy replaces the child's one.
To ensure that the probability of choosing a parent is well-defined, payoffs must always be positive. Since we have assumed $S \in [-1, 1]$ for the payoff matrix (Eq.~\eqref{eq:payoff_matrix}), $S$ may be negative. Therefore, we alter all entries of the payoff matrix by adding to them a constant $\varepsilon > 1$ such that they are all positive. This alteration preserves the type of dilemma. Therefore, we use
\begin{equation}
\label{eq:rescaled_reduced_payoff_matrix}
\mathcal{A}^{\prime\prime} = 
\begin{array}{ccc}
          & \mbox{ } \!\!\!\!{\rm C} & \!\!\!\!\!\!\! {\rm D} \\
          \begin{array}{c} {\rm C} \\ {\rm D} \end{array} & \!\!\!\!\!\!\left(\begin{array}{c} R^\prime \\ T^\prime
          \end{array}\right. & \left.\!\!\!\!\begin{array}{c} S^\prime \\ P^\prime
          \end{array}\right)\\
          & \mbox{ }  & \mbox{ }%
\end{array}
\;,
\end{equation}
where $R^\prime = 1+\varepsilon$, $S^\prime = S+\varepsilon$, $T^\prime = T+\varepsilon$, $P^\prime = \varepsilon$, and $\varepsilon > 1$. For the payoff matrix given by Eq.~\eqref{eq:payoff_matrix}, the payoff of a cooperator is
\begin{align}
\label{eq:payoff_coop_epsilon}
\nonumber \pic^\prime &= \dfrac{(\fc + \fz) \, R^\prime + (1 - \fc) \, S^\prime}{1 + \fz} \\
%
%
     &= \dfrac{(\fc + \fz) \, R + (1 - \fc) \, S}{1 + \fz} + \dfrac{(1 + \fz) \, \varepsilon}{1 + \fz} = \pic + \varepsilon \,.
\intertext{Similarly, we get}
\pid^\prime &= \dfrac{(\fc + \fz) \, T^\prime + (1 - \fc) \, P^\prime}{1 + \fz}  = \pid + \varepsilon \,,
\label{eq:payoff_def_epsilon} 
\end{align}
where $\pic$ and $\pid$ are the payoffs computed using the entries of the reduced payoff matrix (Eq.~\eqref{eq:payoff_matrix}).

One way to introduce selection pressure in the replicator dynamics is to re-formulate the payoffs as follows
\begin{align}
\pic^\prime &= 1 - w + w \, \pic \,,
\label{eq:def_payoff_coop_replicator} \\
\pid^\prime &= 1 - w + w \, \pid \,,
\label{eq:def_payoff_def_replicator}
\end{align}
where $w \in [0,1]$ determines the intensity of selection. The payoffs displayed in Eqs.~\eqref{eq:payoff_coop_epsilon} and \eqref{eq:payoff_def_epsilon}, and those in Eqs.~\eqref{eq:def_payoff_coop_replicator} and \eqref{eq:def_payoff_def_replicator} are equivalent with the identification $w=1/(1+\varepsilon)$. Then, $\pic^\prime$ and $\pid^\prime$ shown in Eqs.~\eqref{eq:payoff_coop_epsilon} and \eqref{eq:payoff_def_epsilon} are both $(1+\varepsilon)$ -times larger (corresponding to a change in the time scale without changing other aspects of replicator dynamics) than $\pic^\prime$ and $\pid^\prime$ given by Eqs.~\eqref{eq:def_payoff_coop_replicator} and \eqref{eq:def_payoff_def_replicator}. Since the condition $\varepsilon > 1$ must hold to ensure that the Moran process is well-defined, one must have $w \in \, \left] \, 0, 1/2 \,\right[$. Using $w$ as a parameter to control the selection pressure and Eq.~\eqref{eq:payoff_matrix} as the payoff matrix, the payoffs are given by
\begin{align}
\pic &= 1 - w + \dfrac{w \bigl[ (\fc + \fz) \, R + (1 - \fc) \, S \bigr]}{1 + \fz}\,, 
\label{eq:payoff_coop_replicator} \\
\pid &= 1 - w + \dfrac{w \bigl[ (\fc + \fz) \, T + (1 - \fc) \, P \bigr]}{1 + \fz}\,,
\label{eq:payoff_def_replicator} \\
\avg{\pi} &=  (\fc + \fz) \, \pic + (1 - \fc) \, \pid \,,
\label{eq:avg_payoff_replicator}
\end{align}
where $\avg{\pi}$ is the payoff averaged over all agents including zealots. The probability of choosing a cooperator or a defector as parent is given by
\begin{align}
\label{eq:prob_parent_replicator_1}
\mathcal{P}_{\rm C} &=  \dfrac{(\fc + \fz) \, \pic}{\avg{\pi}}
\intertext{and}
\label{eq:prob_parent_replicator_2}
\mathcal{P}_{\rm D} &=  \dfrac{(1 - \fc) \, \pid}{\avg{\pi}}\,,
\end{align}
respectively. The replicator dynamics is driven by the difference between the rate at which defectors become cooperators, $\mathbb{T}^+$, and the rate at which cooperators become defectors, $\mathbb{T}^-$. Thus,
\begin{equation}
\label{eq:var_fc_dot_inf_pop_replicator}
\dot{\fc} = \mathbb{T}^+ - \mathbb{T}^- = (1 - \fc) \, \mathcal{P}_{\rm C}  - \fc \, \mathcal{P}_{\rm D} \;.
\end{equation}
The first term on the right-hand side of Eq.~\eqref{eq:var_fc_dot_inf_pop_replicator} represents the probability that a defector is chosen as child and a cooperator as parent. The second term represents the probability that a cooperator is chosen as child and a defector as parent. By substituting Eqs.~\eqref{eq:prob_parent_replicator_1} and \eqref{eq:prob_parent_replicator_2} into Eq.~\eqref{eq:var_fc_dot_inf_pop_replicator}, one obtains
\begin{equation}
\label{eq:var_fc_dot_inf_pop_replicator_2}
\dot{\fc} = \dfrac{(1 - \fc)}{\avg{\pi}} \Bigl[ (\fc + \fz) \, \pic - \fc \, \pid \Bigr] \,.
\end{equation}
%
%
%
\begin{figure*}[ht!]
\centering
\includegraphics[width=0.95\textwidth]{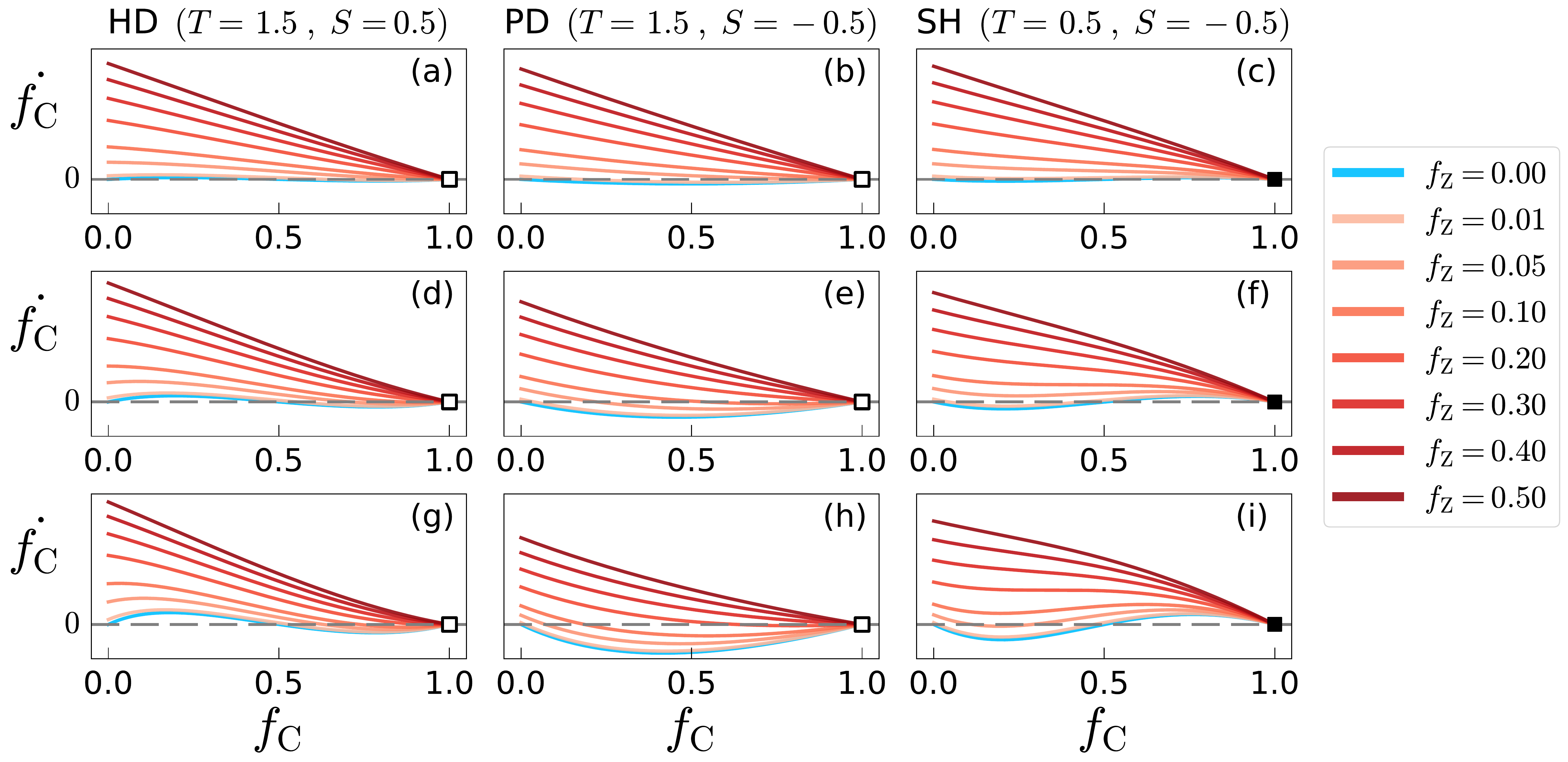}
\caption{Phase portraits of the cooperators' fraction, $\fc$, under the replicator dynamics. Each row corresponds to a different value of selection pressure, $w$, namely $w = 0.1$ (top), $w = 0.3$ (middle), and $w = 0.49$ (bottom). See the caption or Fig.~\ref{fig:results_inf_pop_mf_all_games} for notations and legends.}
\label{fig:mf_replicator_all_w}
\end{figure*}
By substituting Eqs.~\eqref{eq:payoff_coop_replicator}--\eqref{eq:avg_payoff_replicator} into Eq.~\eqref{eq:var_fc_dot_inf_pop_replicator_2}, we obtain
\begin{widetext}
\begin{multline}
\label{eq:evol_coop_replicator}
\dot{\fc} = \cfrac{\bigl(1 - \fc \bigr) \bigl(\fc + \fz \bigr) \left\{ 1 - w + \cfrac{w \bigl[ (\fc + \fz) \, R + (1 - \fc) \, S \bigr]}{1 + \fz} \right\}}{\bigl(\fc + \fz \bigr) \left\{ 1 - w + \cfrac{w \bigl[ (\fc + \fz) \, R + (1 - \fc) \, S \bigr]}{1 + \fz} \right\} - \fc \left\{ 1 - w + \cfrac{w \bigl[ (\fc + \fz) \, T + (1 - \fc) \, P \bigr]}{1 + \fz} \right\}} \\
- \cfrac{\fc \bigl(1 - \fc \bigr) \left\{ 1 - w + \cfrac{w \bigl[ (\fc + \fz) \, T + (1 - \fc) \, P \bigr]}{1 + \fz} \right\}}{\bigl(\fc + \fz \bigr) \left\{ 1 - w + \cfrac{w \bigl[ (\fc + \fz) \, R + (1 - \fc) \, S \bigr]}{1 + \fz} \right\} - \fc \left\{ 1 - w + \cfrac{w \bigl[ (\fc + \fz) \, T + (1 - \fc) \, P \bigr]}{1 + \fz} \right\}} \,.
\end{multline}
\end{widetext}
In Fig.~\ref{fig:mf_replicator_all_w}, we show $\dot{\fc}$ versus $\fc$ for different values of $\fz$ in the case of the HD, PD, and SH games for three values of selection pressure, $w \in \{ 0.1 , 0.3, 0.49 \}$. We observe the presence of a critical mass effect in all the dilemmas regardless of $w$.

One obtains the equilibria of the dynamics given by Eq.~\eqref{eq:evol_coop_replicator} by solving
\begin{widetext}
\begin{multline}
\label{eq:sol_evo_coop_replicator}
\bigl(1 - \fc \bigr) \Biggl\{\!\!\!\Biggl\{ \bigl(\fc + \fz \bigr) \left\{ 1 - w + \cfrac{w \bigl[ (\fc + \fz) \, R + (1 - \fc) \, S \bigr]}{1 + \fz} \right\}  \Biggr.\Biggr. \\
- \Biggl.\Biggl. \fc \left\{ 1 - w + \cfrac{w \bigl[ (\fc + \fz) \, T + (1 - \fc) \, P \bigr]}{1 + \fz} \right\} \Biggr\}\!\!\!\Biggr\} = 0 \,.
\end{multline}
\end{widetext}
The solution $\fc = 1$ corresponds to full cooperation. As expected, this equilibrium exists regardless of the value of, $\fz$, $w$, and the type of the game. The other solutions of Eq.~\eqref{eq:sol_evo_coop_replicator} are the solutions of the following equation:
\begin{widetext}
\begin{equation}
\label{eq:second_eq_replicator}
\bigl(\fc + \fz \bigr) \left\{ 1 - w + \cfrac{w \bigl[ (\fc + \fz) \, R + (1 - \fc) \, S \bigr]}{1 + \fz} \right\}
- \fc \left\{ 1 - w + \cfrac{w \bigl[ (\fc + \fz) \, T + (1 - \fc) \, P \bigr]}{1 + \fz} \right\} = 0 \,.
\end{equation}
\end{widetext}
Solving Eq.~\eqref{eq:second_eq_replicator} is not straightforward. However, we can extract some useful information under specific conditions. First, we verify whether $\fc = 0$ is a solution of Eq.~\eqref{eq:second_eq_replicator} or not. By setting $\fc = 0$, one obtains
\begin{equation}
\label{eq:solutions_infpop_nozeal_replicator}
\fz  \left\{ 1 - w + \dfrac{w \bigl[ \fz \, R + S \bigr]}{1 + \fz} \right\} = 0 \,.
\end{equation}
Equation \eqref{eq:solutions_infpop_nozeal_replicator} has two solutions, \ie
\begin{gather}
\label{eq:solutions_system_inf_pop_replicator_1}
\fz = 0
\intertext{and}
\label{eq:solutions_system_inf_pop_replicator_2}
1 - w + \dfrac{w \bigl[ \fz \, R + S \bigr]}{1 + \fz} = 0 \,.
\end{gather}
The solution corresponding to Eq.~\eqref{eq:solutions_system_inf_pop_replicator_1} tells us that $\fc = 0$ is a solution of Eq.~\eqref{eq:sol_evo_coop_replicator} only in the absence of zealots, which is the same as the case of the Fermi rule. Next, Eq.~\eqref{eq:solutions_system_inf_pop_replicator_2} is equivalent to
\begin{equation}
\label{eq:condition_2_replicator}
\fz = \dfrac{1 + w \left( S -1 \right)}{w \left( 1 - R \right) - 1}\,.
\end{equation}
The denominator is always negative. Therefore, $\fz \geq 0$ requires that the numerator must be nonpositive, giving
\begin{equation}
\label{eq:cond_w_s}
w \leq \dfrac{1}{1 - S}\,.
\end{equation}
Inequality \eqref{eq:cond_w_s} is always satisfied for $S \in [ -1, 1 [$, which means that there exists a value $\fz \geq 0$ such that there exists a solution $0 < {\fc}^\star < 1$ of Eq.~\eqref{eq:sol_evo_coop_replicator} regardless of $w$.

Let us study now the effects of selection pressure, $w$, on the solutions of Eq.~\eqref{eq:second_eq_replicator}. In the zero selection limit, Eq.~\eqref{eq:second_eq_replicator} becomes
\begin{equation}
\fc + \fz = \fc  \qquad \Longleftrightarrow \qquad \fz = 0 \,.
\end{equation}
Hence, in the zero selection limit, the only solutions to $\dot{\fc} = 0$ apart from ${\fc}^\star = 1$ are those obtained in the absence of zealots. The strong selection limit, $w = 1$, returns a different scenario. By setting $w = 1$ in Eq.~\eqref{eq:second_eq_replicator} one obtains
\begin{widetext}
\begin{equation}
\bigl(\fc + \fz \bigr) \dfrac{\bigl[ (\fc + \fz) \, R + (1 - \fc) \, S \bigr]}{1 + \fz} =
\fc \dfrac{\bigl[ (\fc + \fz) \, T + (1 - \fc) \, P \bigr]}{1 + \fz} \,,
\end{equation}
which leads to
\begin{equation}
\label{eq:second_order_replicator}
\underbracket{\bigl( R - S - T + P \bigr)}_{A} {\fc}^2 + \underbracket{\Bigl[ 2\fz R + (1-\fz) S - \fz T - P \Bigr]}_{B} \fc + \underbracket{\left( {\fz}^2 R + \fz S \right)}_{C} = 0 \,.
\end{equation}
\end{widetext}
The solutions of Eq.~\eqref{eq:second_order_replicator} are ${\fc}^\star = \tfrac{-B \pm \sqrt{B^2 - 4AC}}{2A}$. Therefore, there are at most two other solutions to Eq.~\eqref{eq:sol_evo_coop_replicator}. There is no real solution ${\fc}^\star$ if $B^2 - 4AC < 0$, which is the case if and only if $\fz >\tilde{\fz}$, if such an $\tilde{\fz}$ exists. The value $\tilde{\fz}$ corresponds to the critical mass of zealots needed to observe only full cooperation. The values of $\tilde{\fz}$ are displayed in Table~\ref{tab:replicator_crit_thres_fz}.

%
%
%
  \begin{table}[h]
  {%
  \def\arraystretch{1.35}
  \newcommand{\mc}[3]{\multicolumn{#1}{#2}{#3}}
  \definecolor{tcA}{rgb}{1,0.678431,0.305882}
  \definecolor{Gray}{gray}{0.8}
  \newcolumntype{a}{>{\columncolor{Gray}}c}
  \begin{center}
  \begin{large}
  \begin{tabular}[c]{|l|c|c|c|}
  \cline{2-4}
  \mc{1}{c|}{$\,$} & \textbf{HD} & \textbf{PD} & \textbf{SH}\\ \hline
  $R$ & \mc{3}{c|}{1.0} \\ \hline
  $P$ & \mc{3}{c|}{0.0} \\ \hline
  $T$ & $\tfrac{3}{2}$ & $\tfrac{3}{2}$ & $\tfrac{1}{2}$ \\ \hline
  $S$ & $\tfrac{1}{2}$ & $-\tfrac{1}{2}$ & $-\tfrac{1}{2}$\\ \hline
  $\varepsilon$ & \mc{3}{c|}{1.0408} \\ \hline\hline
  \rowcolor{Gray}
  $\tilde{\fz}$ & {\small N/A} & $0.5$ & 0.06 \\ \hline 
  \end{tabular}
  \end{large}
  \end{center}
  \caption{Minimum fraction of zealots, $\tilde{\fz}$, required to observe exclusively the full cooperative equilibrium (\ie{} $\fc = 1$) for a replicator dynamics in the strong selection limit (\ie{} $w = 0.49$).}
  \label{tab:replicator_crit_thres_fz}
  }%
  \end{table}
%


\section{Numerical results in mean-field populations}
\label{sec:mean_field-numerics}

\subsection{Simulation setup}
\label{ssec:simu_setup}

We provide here some details about the numerical implementation of our agent-based simulations. Given a population of $N$ agents, we initialize the agents' strategies such that we have a fraction $\fz$ of zealots. Then, we start the evolutionary dynamics where, at each time step, we first perform a uniform random sampling of $N_g = 5N$ pairs of agents and let them play and accumulate payoff according to Eq.~\eqref{eq:payoff_matrix}. In this way, each agent has played ten times on average. Setting $N_g = 5N$ yields a good trade-off between having a sufficiently many games per round played per agent, and keeping the overall computation time reasonably short. After agents accumulate their payoff, they update their strategies, payoffs are reset to zero, and we measure the fraction of cooperators among normal agents, $\fc$, and begin a new round. We let the dynamics to evolve for a transient of $t_{\text{tr}}$ time steps during which we only store the value of $\fc$. Then, for $t > t_{\text{tr}}$ we evaluate the average, and standard deviation, of $\fc$ over a moving window of length $t_{\text{tr}}$. For PD and SH games we expect that the dynamics ends in one of its two adsorbing states: $\fc = 1$ and $\fc = 0$, respectively. However, to expedite the computation, instead of waiting until the system reaches the adsorbing state, we let the evolution stop once $\avg{\fc} > 0.95$ (or $\avg{\fc} < 0.05$) and $\sigma_{\fc} < 0.01$ because we assume that the system is then in a metastable state, and that it will reach the adsorbing state for sure. In the case of the HD game, we know that the stationary state is $0 < \fc < 1$ and that it takes a lot of time to reach the stationariety. In such case (or if the dynamics does not end up in any of the aforementioned conditions), the dynamics stops after $t_{\max} = 5\cdot10^5$ steps. Then, we compute $\avg{\fc}$ over the last $t_{\text{tr}}$ steps and use such value as the ``final'' value of cooperation. We remark that the aforementioned approach corresponds to computing the intermediate equilibrium point of the mean-field dynamics.
%
%
%
\begin{figure*}[t!]
\centering
\includegraphics[width=0.98\textwidth]{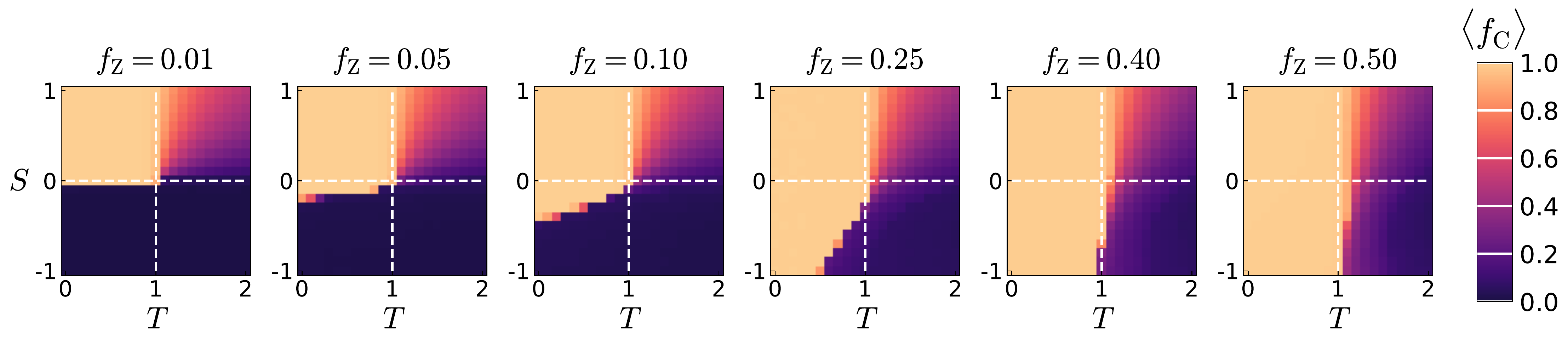}
\caption{Average fraction of cooperation among normal agents, $\avg{\fc}$, for the considered $(T,S)$ parameter region using the Fermi update rule with $\beta = 10$. Each panel corresponds to a different value of $\fz$. Results are averaged over $N_{\text{rep}} = 50$ realizations over a population of $N = 1000$ agents.}
\label{fig:mf_numeric_coop}
\end{figure*}
%

%
%
\begin{figure*}[ht!]
\centering
\includegraphics[width=0.98\textwidth]{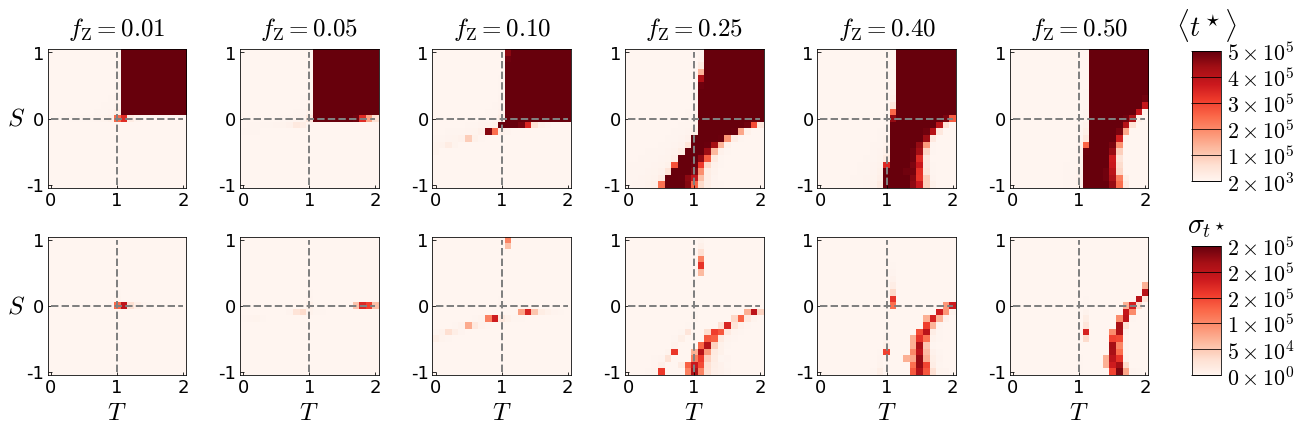}
\caption{(top) Average time needed to reach the stationary state, $\avg{t^\star}$, and (bottom) its standard deviation, $\sigma_{t^\star}$, in the $(T, S)$ parameter space used in Fig.~\ref{fig:mf_numeric_coop} using the Fermi update rule with $\beta = 10$. Each panel corresponds to a different value of $\fz$. Results are averaged over $N_{\text{rep}} = 50$ realizations over a population of $N = 1000$ agents.}
\label{fig:mf_numeric_conv_time}
\end{figure*}

\subsection{Exploration of the $(T,S)$ space}
\label{ssec:ts_exploration}

In the main text, we focused on typical $(T, S)$ values. To examine the generality of our results in terms of $T$ and $S$, we numerically explore the $(T,S)$ space in this section. We set $N = 1000$ and $\beta=10$. In Fig.~\ref{fig:mf_numeric_coop} we show the average fraction of cooperation among normal agents, $\avg{\fc}$, as we vary $T$, $S$, and the fraction of zealots, $\fz$ for the Fermi update rule. The figure indicates that, for the SH game, the transition from the equilibrium with little cooperation to that with full cooperation occurs suddenly as one increases $\fz$. Note that this is the case for the entire parameter $(T, S)$ region corresponding to the SH game (\ie{} $T < 1$ and $S < 0$). In contrast, for the HD game, the effect of $\fz$ is only little and gradual in the entire parameter region (\ie{} $T > 1$ and $S > 0$). The results for the PD game (\ie{} $T > 1$ and $S < 0$) are mixed. In Fig.~\ref{fig:mf_numeric_conv_time}, we display the time needed for the simulation to converge to the stationary state, $\avg{t^\star}$, and its standard deviation, $\sigma_{t^\star}$.

\pagebreak

\section{Numerical results for networks}
\label{sec:networks-numerics}

%
%
\begin{figure*}[h!]
\centering
\includegraphics[width=0.98\textwidth]{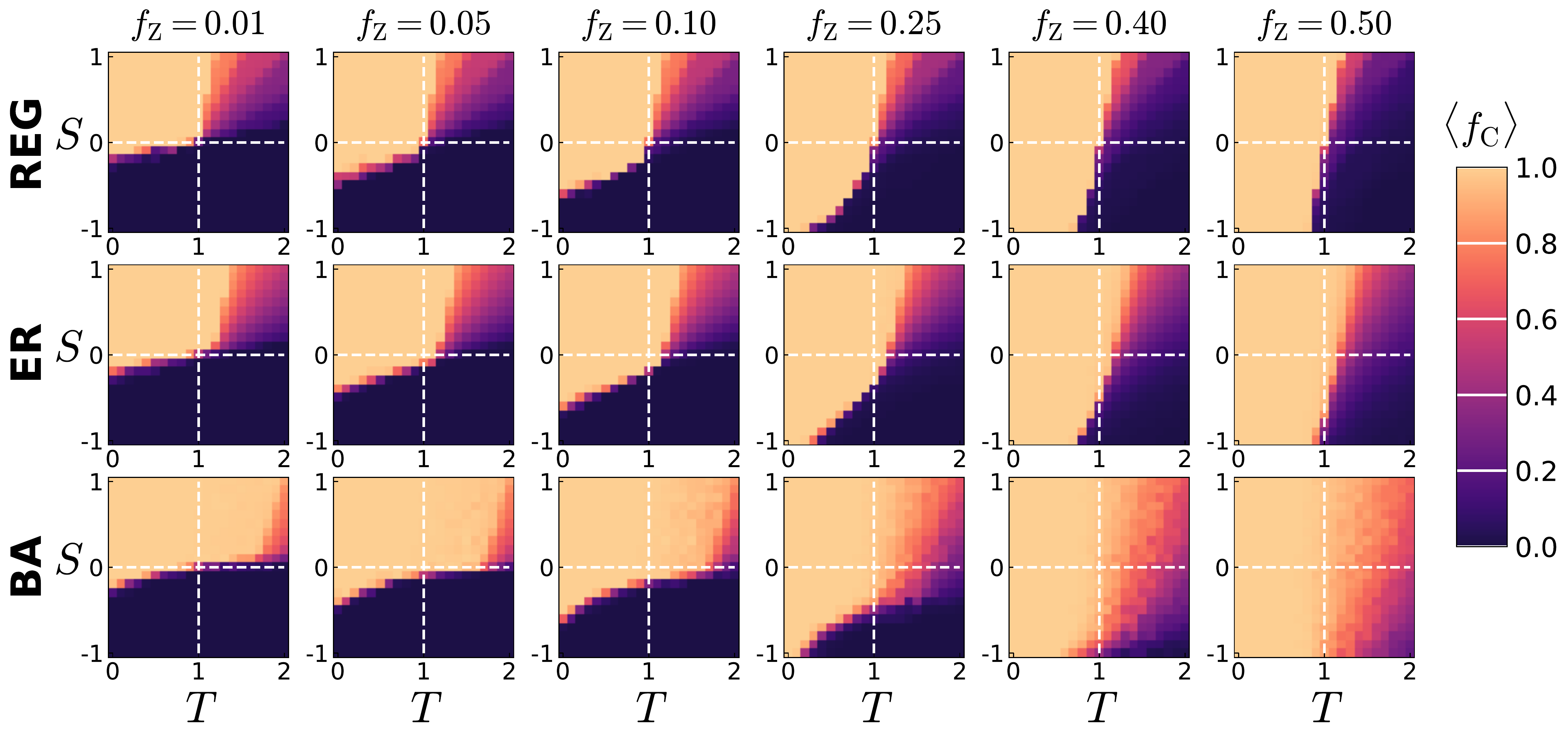}
\caption{Average fraction of cooperators among normal agents, $\avg{\fc}$, in the $(T,S)$ space for different networks in the additive payoff scheme. The first, second, and third rows correspond to random regular lattice (REG), Erd\H{o}s-Reny\'i (ER) networks, and B\'arabasi-Albert (BA) networks, respectively. Each column accounts for a different fraction of zealots, $\fz$. We set $N=1000$ nodes and $\avg{k}=6$. We use the Fermi update rule with $\beta = 10$. Results are averaged over $N_{\text{rep}} = 50$ different realizations.}
\label{fig:all_nets_coop_nodegq}
\end{figure*}
%
%
%
%
\begin{figure*}[h!]
\centering
\includegraphics[width=0.98\textwidth]{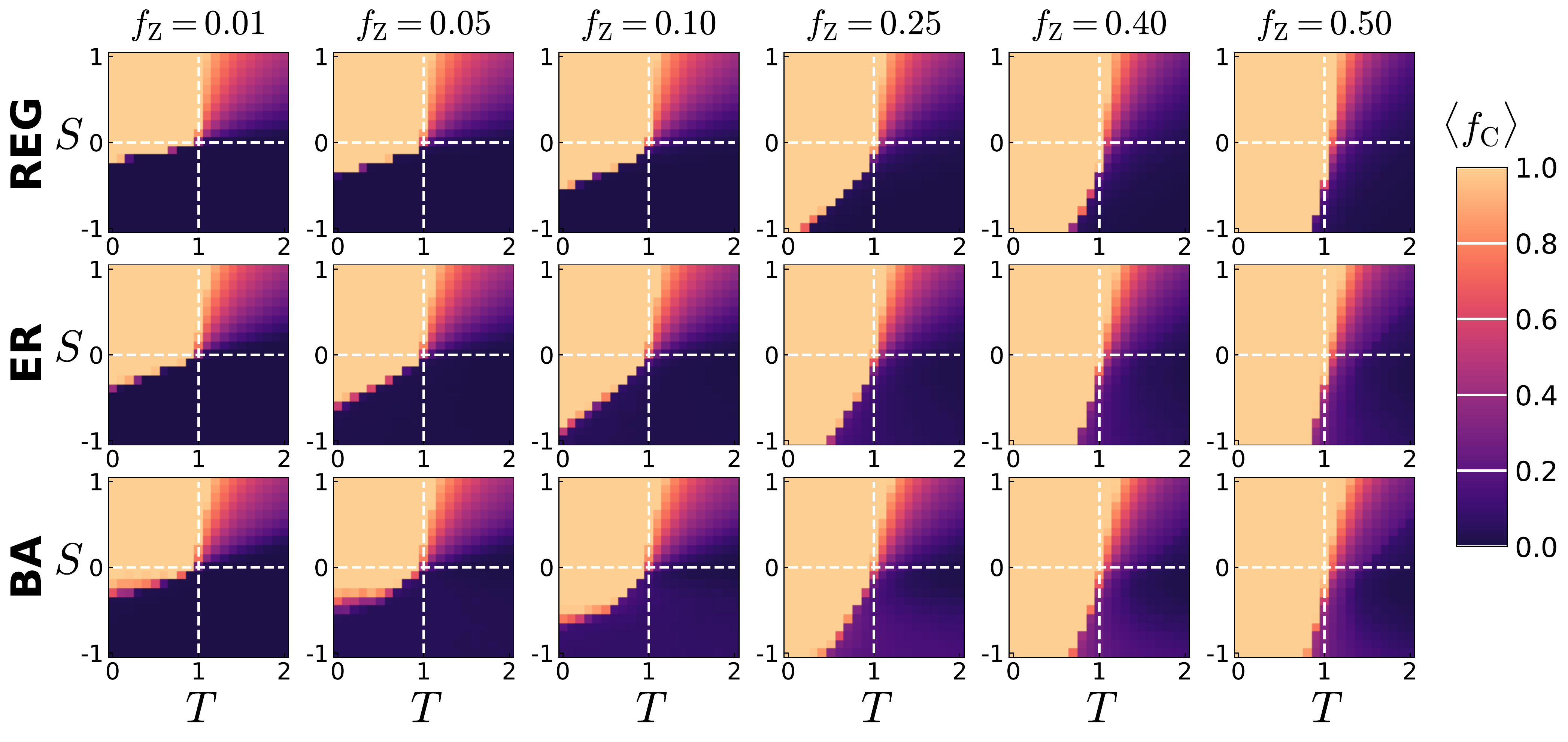}
\caption{Fraction of cooperators among normal agents, $\avg{\fc}$, for networks in the $(T,S)$ space for the average payoff scheme. The first, second, and third rows correspond to random regular lattice (REG), Erd\H{o}s-Reny\'i (ER) networks, and B\'arabasi-Albert (BA) networks, respectively. Each column accounts for a different fraction of zealots, $\fz$. Each network has $N=1000$ nodes and $\avg{k}=6$. We use the Fermi update rule with $\beta = 10$. Results are averaged over $N_{\text{rep}} = 50$ different realizations.}
\label{fig:all_nets_coop_degq}
\end{figure*}

We have also tested the effect of zealotry on networks for various values of $T$ and $S$. The degree distribution is a delta function for REG networks, a Poisson distribution for ER networks, and a power law for BA networks \cite{latora-book-2017}. Moreover, to exclude any other effect than those associated with degree heterogeneity, we fix the number of nodes, $N = 1000$, and the average degree, $\avg{k} = 6$, to be the same for the three types of networks. It is worth noting that degree heterogeneity may boost cooperation \cite{santos-prl-2005,santos-pnas-2006}, or decrease it when a cost is associated with each interaction \cite{masuda-proc_roy_soc_b-2007}. For this reason, we consider two scenarios: one in which we divide the payoff accumulated by each agent by its degree (\ie{} number of connections) to suppress the effect of degree heterogeneity, and another in which we do not carry out the division. The former approach is known as average payoff scheme \cite{tomassini-int_jour_mod_phys_c-2007}, while the other is usually called additive payoff scheme \cite{santos-prl-2005}.

%
%
%
%
%

Figure \ref{fig:all_nets_coop_nodegq} shows the fraction of cooperators for the different networks under the additive payoff scheme. A tiny fraction of zealots is enough to spur almost complete cooperation in the HD game when played on a BA network. However, in agreement with the results for the well-mixed populations, an excessive fraction of zealots decreases the fraction of normal cooperators, making defection more appealing as shown by the re-emergence of defection in the upper right corner of the $(T,S)$ space. Moreover, BA networks considerably increase the cooperation in the PD game, albeit such phenomenon occurs only when $\fz > 0.25$. The SH game, instead, does not display any remarkable difference with the phenomenology observed in well-mixed populations.

%
%

In contrast to the case of additive payoff scheme, under the average payoff scheme the amount of cooperation over the $(T,S)$ space does not remarkably depend on the network structure. Figure \ref{fig:all_nets_coop_degq} shows the average fraction of cooperators, $\avg{\fc}$, for different values of $T$ and $S$, different network structures, and the average payoff scheme. As expected, the pattern of cooperation is roughly independent of the network structure, although ER and BA networks slightly promote cooperation compared to the well-mixed populations in an extensive part of the $(T, S)$ space corresponding to the PD game for $\fz \geq 0.25$.

\pagebreak

%
%


\begin{thebibliography}{99}

\bibitem{eckmann-rev_mod_phys-1985} J.-P. Eckmann, and D. Ruelle, Ergodic theory of chaos and strange attractors. Rev. Mod. Phys. {\bf 57}, 617--656 (1985). 

\bibitem{buldyrev-nature-2010} S.~V. Buldyrev, R. Parshani, G. Paul, H.~E. Stanley, and S. Havlin, Catastrophic cascade of failures in interdependent networks. Nature, {\bf 464}, 1025--1028 (2010). 

\bibitem{gleeson-prx-2016} J.~P. Gleeson, K.~P. O'Sullivan, R. \'Alvarez Ba\~nos, and Y. Moreno, Effects of network structure, competition and memory time on social spreading phenomena. Phys. Rev. X, {\bf 6}, 021019 (2016). 

\bibitem{pound-j_am_chem_soc-1952} G.~M. Pound, and V.~K. La Mer. Kinetics of crystalline nucleus formation in supercooled liquid Tin$^{1,2}$. J. Am. Chem. Soc., {\bf 74} (9), 2323--2332 (1952). 


\bibitem{achlioptas-science-2009} D. Achlioptas, R.~M. D'Souza, and J. Spencer. Explosive percolation in random networks. Science, {\bf 323}, 1453--1455 (2009). 


\bibitem{d_souza-adv_phys-2019} R.~M. D'Souza, J. G\'omez-Garde\~nes, J. Nagler, and A. Arenas. Explosive phenomena in complex networks. Adv. Phys., {\bf 68}, 123--223 (2019). 

\bibitem{kori-prl-2004} H. Kori, and A.~S. Mikhailov, Entrainment of randomly coupled oscillator networks by a pacemaker. Phys. Rev. Lett., {\bf 93}, 254101 (2004). 

\bibitem{brocard-neuron-2013} F. Brocard, N.~A. Shevtsova, M. Bouhadfane, S. Tazerart, U. Heinemann, I.~A. Rybak, and L. Vinay, Activity-dependent changes in extracellular Ca2+ and K+ reveal pacemakers in the spinal locomotor-related network. Neuron {\bf 77}, 1047--1054 (2013). 

\bibitem{yllanes-njp-2017} D. Yllanes, M. Leoni, and M.~C. Marchetti, How many dissenters does it take to disorder a flock? New J. Phys., {\bf 19}, 103026 (2017). 

\bibitem{juul-pre-2019} J.~S. Juul and M.~A. Porter, Hipsters on networks: How a minority group of individuals can lead to an antiestablishment majority. Phys. Rev. E, {\bf 99}, 022313 (2019). 

\bibitem{young-j_econ_soc-1993} H. P. Young, The evolution of conventions. Econometrica, {\bf 61}, 57--84 (1993). 

\bibitem{baronchelli-r_soc_op_sci-2018} A. Baronchelli, The emergence of consensus: A primer. Roy. Soc. Open Sci., {\bf 5}, 172189 (2018). 

\bibitem{galam-phys_a-2007} S. Galam and F. Jacobs, The role of inflexible minorities in the breaking of democratic opinion dynamics. Physica A {\bf 381} 366--376 (2007). 

\bibitem{tyson-j_am_hist-1998} T.~B. Tyson, Robert F. Williams, ``Black Power,'' and the roots of the african american freedom struggle. J. Am. Hist., {\bf 85}, 540--570 (1998). 

\bibitem{opp-am_soc_rev-1993} K.-D. Opp and C. Gern. Dissident groups, personal networks, and spontaneous cooperation: the east german revolution of 1989. Am. Sociol. Rev. {\bf 58}, 659--680 (1993). 

\bibitem{moss_kanter-am_jour_soc-1977} R. Moss Kanter, Some effects of proportions on group life: skewed sex ratios and responses to token women. Am. Jour. Sociol. {\bf 82}, 965--990 (1977). 

\bibitem{rochmyaningsih-science-2018} D. Rochmyaningsih, Indonesian fatwa causes immunization rates to drop. Science, {\bf 362}, 628--629 (2018). 

\bibitem{couzin-science-2011} I.~D. Couzin, C.~C. Ioannou, G. Demirel, T. Gross, C.~J. Torney, A. Hartnett, L. Conradt, S.~A. Levin, and N.~E. Leonard, Uninformed individuals promote democratic consensus in animal groups. Science, {\bf 334}, 1578--1580 (2011). 

\bibitem{feinerman-nphys-2018} O. Feinerman, I. Pinkoviezky, A. Gelblum, E. Fonio, and N.~S. Gov, The physics of cooperative transport in groups of ants. Nat. Phys., {\bf 14}, 683 (2018). 

\bibitem{dyer-animal_beh-2008} J.~R.~G. Dyer, C.~C. Ioannou, L.~J. Morrell, D.~P. Croft, I.~D. Couzin, D.~A. Waters, and J.~Krause. Consensus decision making in human crowds, Anim. Behav., {\bf 75}, 461--470 (2008). 


\bibitem{mobilia-prl-2003} M. Mobilia, Does a single zealot affect an infinite group of voters? Phys. Rev. Lett. {\bf 91}, 028701 (2003). 


\bibitem{xie-pre-2011} J. Xie, S. Sreenivasan, G. Korniss, W. Zhang, C. Lim, and B.~K. Szymanski, Social consensus through the influence of committed minorities. Phys. Rev. E, {\bf 84}, 011130 (2011). 

\bibitem{mistry-pre-2015} D. Mistry, Q. Zhang, N. Perra, and A. Baronchelli, Committed activists and the reshaping of status-quo social consensus. Phys. Rev. E {\bf 92}, 042805 (2015). 

\bibitem{waagen-pre-2015} A. Waagen, G. Verma, K. Chan, A. Swami, and R. D'Souza, Effect of zealotry in high-dimensional opinion dynamics models. Phys. Rev. E, {\bf 91}, 022811 (2015). 

\bibitem{rodriguez-pone-2016} N. Rodriguez, J. Bollen, and Y.-Y. Ahn, Collective dynamics of belief evolution under cognitive coherence and social conformity. PLoS ONE {\bf 11}, e0165910 (2016). 

\bibitem{turalska-scirep-2013} M. Turalska, B.~J. West, and P. Grigolini, Role of committed minorities in times of crisis. Sci. Rep., {\bf 3}, 1371 (2013). 

\bibitem{masuda-scirep-2012} N. Masuda, Evolution of cooperation driven by zealots. Sci. Rep., {\bf 2}, 646 (2012). 

\bibitem{mobilia-pre-2012} M. Mobilia, Stochastic dynamics of the prisoner's dilemma with cooperation facilitators. Phys. Rev. E, {\bf 86}, 011134 (2012). 

\bibitem{nakajima-math_bio-2015} Y. Nakajima and N. Masuda, Evolutionary dynamics in finite populations with zealots. J. Math. Biol., {\bf 70}, 465--484 (2015). 

\bibitem{jensen-prl-2018} P. Jensen, T. Matreux, J. Cambe, H. Larralde, and E. Bertin, Giant catalytic effect of altruists in Schelling's segregation model. Phys. Rev. Lett. {\bf 120}, 208301 (2018). 

\bibitem{mobilia-jstat-2007} M. Mobilia, A. Petersen, and S. Redner, On the role of zealotry in the voter model. Jour. Stat. Mech.: Theo. \& Exper., {\bf 2007}, P08029 (2007). 

\bibitem{centola-science-2018} D. Centola, J. Becker, D. Brackbill, and A. Baronchelli, Experimental evidence for tipping points in social convention. Science, {\bf 360}, 1116--1119 (2018). 

\bibitem{gintis-book-2009} H. Gintis, \emph{Game Theory Evolving: A Problem-Centered Introduction to Modeling Strategic Interaction} (2nd ed.) (Princeton University Press, Princeton, 2009).

\bibitem{santos-pnas-2006} F. Santos, J. Pacheco, and T. Lenaerts, Evolutionary dynamics of social dilemmas in structured heterogeneous populations. Proc. Nat. Acad. Sci. USA, {\bf 103}, 3490--3494 (2006). 

\bibitem{santos-prl-2005} F. Santos, and J. Pacheco, Scale-free networks provide a unifying framework for the emergence of cooperation. Phys. Rev. Lett., {\bf 95}, 098104 (2005). 

\bibitem{blume-gam_eco_beh-1993} L.~E. Blume, The statistical mechanics of strategic interaction. Games Econ. Behav., {\bf 5}, 387--424 (1993). 

\bibitem{szabo-pre-1998} G. Szab\'o, and C. T\H{o}ke, Evolutionary prisoner's dilemma game on a square lattice. Phys. Rev. E, {\bf 58}, 69--73 (1998). 

\bibitem{nowak-book-2006}  M.~A. Nowak, \emph{Evolutionary Dynamics -- Exploring the Equations of Life}. (Belknap Press, Cambdrige, 2006).

\bibitem{matsuzawa-pre-2016} R. Matsuzawa, J. Tanimoto, and E. Fukuda, Spatial prisoner's dilemma games with zealous cooperators. Phys. Rev. E, {\bf 94}, 022114 (2016). 

\bibitem{nowak-science-2006} M.~A. Nowak, Five rules for the evolution of cooperation. Science, {\bf 314}, 1560--1563 (2006). 

\bibitem{szabo-phys_rep-2007} G. Szab\'o and G. F\'ath, Evolutionary games on graphs. Phys. Rep., {\bf 446}, 97--216 (2007).

\bibitem{erdos-paper-1959} P. Erd\H{o}s and A. R\'enyi, On random graphs. Publicationes Mathematicae (Debrecen) {\bf 6}, 290 (1959).

\bibitem{barabasi-science-1999} A.-L. Barab\`asi and R. Albert, Emergence of scaling in random networks. Science, {\bf 286}, 509--512 (1999). 

\bibitem{latora-book-2017} V. Latora, V. Nicosia, and G. Russo, \emph{Complex Networks}. (Cambridge University Press, Cambridge, UK, 2017).  

\bibitem{roca-phys_lif_rev-2009} C.~P. Roca, J.~A. Cuesta, and A. S\'anchez, Evolutionary game theory: temporal and spatial effects beyond replicator dynamics. Phys. Life Rev., {\bf 6}, 208--249 (2009). 

\bibitem{abramson-pre-2001} G. Abramson, and M. Kuperman, Social games in a social network. Phys. Rev. E, {\bf 63}, 030901 (2001). 

\bibitem{grujic-pone-2010} J. Gruji\'c, C. Fosco, L. Araujo, J.A. Cuesta, and A. S\'anchez, A. Social experiments in the mesoscale: humans playing a spatial prisoner's dilemma. PloS ONE, {\bf 5}, e13749 (2010). 

\bibitem{liu-pre-2012} X.-T. Liu, Z.-X. Wu, and L. Zhang, Impact of committed individuals on vaccination behavior. Phys. Rev. E, {\bf 86}, 051132 (2012). 

\bibitem{antonioni-prl-2017} A. Antonioni and A. Cardillo, Coevolution of synchronization and cooperation in costly networked interactions. Phys. Rev. Lett., {\bf 118}, 238301 (2017). 

\bibitem{delellis-ieee-2010} P. DeLellis, M. Bernardo, T. E. Gorochowski, and G. Russo, Synchronization and control of complex networks via contraction, adaptation and evolution. IEEE Circuits and Systems Magazine, {\bf 10}, 64--82. (2010). 


\bibitem{szolnoki-pre-2016} A. Szolnoki and M. Perc, Zealots tame oscillations in the spatial rock-paper-scissors game. Phys. Rev. E, {\bf 93}, 062307 (2016). 

\bibitem{mc_creery-insec_soc-2017} H.~F. McCreery, A comparative approach to cooperative transport in ants: individual persistence correlates with group coordination. Insectes Sociaux, {\bf 64}, 535--547 (2017). 

\bibitem{kabla-j_roy_soc_int-2012} A.~J. Kabla, Collective cell migration: Leadership, invasion and segregation. J. R. Soc. Interface, {\bf 9}, 3268--3278 (2012). 

\bibitem{arendt-comp_math_org_theo-2015} D.~L. Arendt, and L.~M. Blaha, Opinions, influence, and zealotry: A computational study on stubbornness. Comput. Math. Organ. Theory, {\bf 21}, 184 (2015). 

\bibitem{hunter-matplotlib-2007} J.~D. Hunter, Matplotlib: A 2D graphics environment. Computing in Science \& Engineering, {\bf 9}, 90--95, (2007). 

\bibitem{antal-bull_math_bio-2006} T. Antal, and I. Scheuring, Fixation of strategies for an evolutionary game in finite populations. Bull. Math. Biol., {\bf 68}, 1923--1944 (2006). 


\bibitem{strogatz-book-1994} S.~H. Strogatz, \emph{Nonlinear Dynamics and Chaos: With Applications to Physics, Biology, Chemistry and Engineering} (CRC Press, Boca Raton, 1994).

\bibitem{moran-book-1962} P.~A.~P. Moran, \emph{The Statistical Processes of Evolutionary Theory} (Oxford University Press, Oxford, UK, 1962).


\bibitem{masuda-proc_roy_soc_b-2007} N. Masuda, Participation costs dismiss the advantage of heterogeneous networks in evolution of cooperation. Proc. Roy. Soc. B, {\bf 274}, 1815--1821 (2007). 

\bibitem{tomassini-int_jour_mod_phys_c-2007} M. Tomassini, E. Pestelacci, and L. Luthi, Social dilemmas and cooperation in complex networks. Int. J. Mod. Phys. C, {\bf 18}, 1173--1185 (2007). 




%
%
%
%
%
%
%
%
%
%
%
%



\end{thebibliography}
\end{document}